\documentclass{jfm}

\usepackage[intlimits]{amsmath}
\usepackage{amsmath}
\usepackage{amssymb}
\usepackage[usenames, dvipsnames]{color}
\newcommand {\ra}{\rightarrow}

\newcommand {\pt}{\partial}

\def\v{\boldsymbol{v}}
\def\u{\boldsymbol{u}}
\def\k{\boldsymbol{k}}
\def\kp{k_{1\perp}}
\def\kpp{k_{2\perp}}
\def\kppp{k_{3\perp}}
\def\kpppp{k_{4\perp}}
\def\bkp{\boldsymbol{k}_{1\perp}}
\def\bkpp{\boldsymbol{k}_{2\perp}}
\def\bkppp{\boldsymbol{k}_{3\perp}}
\def\bkpppp{\boldsymbol{k}_{4\perp}}

\def\x{\boldsymbol{x}}
\def\y{\boldsymbol{y}}
\def\z{\boldsymbol{z}}
\def\r{\boldsymbol{r}}
\def\brho{\boldsymbol{\rho}}

\def\n{\bnabla}
\def\bn{\bnabla}
\def\bO{\boldsymbol{\Omega}}
\def\O{\Omega}

\def\P{\mathbf{P}}
\let\^\widehat
\def\*{\boldsymbol{\cdot}}
\let\p\partial
\def\H{\mathcal H}
\def\[{\left[}\def\]{\right]}
\def\({\left(}\def\){\right)}
\let\wt\widetilde
\def\REF#1{Eq.~\Ref{#1}}\def\Ref#1{(\ref{#1})}

\begin{document}

\shorttitle{Complete Hamiltonian formalism} 
\shortauthor{A. A. Gelash et al} 

\title{Complete Hamiltonian formalism for inertial waves in rotating fluids}

\author{
A.\,A.~Gelash\aff{1,2}\corresp{\email{agelash@gmail.com}},
V.\,S.~L'vov\aff{3} \and V.\,E.~Zakharov\aff{1,4,5}}

\affiliation{
\aff{1}Novosibirsk State University, Novosibirsk, 630090, Russia
\aff{2}Kutateladze Institute of Thermophysics, SB RAS, Novosibirsk, 630090, Russia
\aff{3}Department of Chemical Physics, The Weizmann Institute of Science, Rehovot, 76100, Israel
\aff{4}Department of Mathematics, University of Arizona, AZ 857201 Tucson, USA
\aff{5}Lebedev Physical Institute, Russian Academy of Sciences, Moscow, 119991, Russia}

\maketitle

\begin{abstract}
Complete Hamiltonian formalism is suggested for inertial waves in rotating incompressible fluid. Resonance three-wave interaction processes -- decay instability and confluence of two waves  -- are shown to play a key role in the weakly nonlinear dynamics and statistics of inertial waves in the rapid rotation case. Future applications of the Hamiltonian approach in inertial wave theory are investigated and discussed.
\end{abstract}

\section{Introduction}

Rotation fundamentally changes the behaviour of fluid flows, which served as the basis of one of the classic chapters in hydrodynamics (see, e.g., the monograph of~\cite{greenspan1968theory}). The presence of the Coriolis force leads to the emergence of a special type of waves, inertial waves, that appear in geophysical, astrophysical and industrial flows. In an incompressible fluid rotating with angular velocity $\bO$, inertial waves have the following dispersion law:
\begin{equation}\label{dispersion}
\omega_{\k} = 2\O|\cos\theta_{\k}| \,.
\end{equation}
Here, $\theta_{\k}$ is the angle between the wave vector $\k$ and the angular velocity vector $\bO$. The linear theory of inertial waves is well known and thoroughly described, for example, in the monographs of~\cite{greenspan1968theory} and~\cite{landau1987fluid}. These waves propagate inside the volume of the fluid and have circular polarization. They have attracted special attention due to their unusual  properties of reflection from the rigid boundary~\citep{maas1997observation}. Namely, the angle of reflection is determined not by the angle of incidence to the boundary, but by the angle with respect to the rotating axes~\citep{greenspan1968theory}. The plane inertial wave is the exact solution of the fully nonlinear Euler equations, which will be discussed in a following section.

Real physical rotating flows can be complicated by the influence of other factors, such as gravitation or electromagnetic forces (in the latter case, the fluid is assumed to be an electric conductor), and the angular velocity of the fluid can depend on the radius. In such cases, the dispersion law is more complicated than the presented relation~(\ref{dispersion})~(see, e.g. \cite{gal2013waves}). Other important effects also appear due to nonlinear wave interactions.

The degree of nonlinearity of fluid flows is characterized by \emph{the nonlinearity parameter} $\delta$, which is the ratio of the nonlinear and linear non-dissipative terms of the Euler equations. In the case of rotating fluids, the nonlinearity parameter (known as the Rossby number) can be written as:
\begin{equation}\label{delta}
  \delta \equiv \frac{\left|\left(\u\times {\rm rot}\,\u\right)\right|}
    {\left|\left(\bO\times \u\right)\right|}
\,.\end{equation}
Here, velocity $\u$ is the characteristic value of the wave amplitude. In a rapidly rotating fluid , $\delta \ll 1$; thus, the inertial waves interact weakly. In this case, we can apply perturbation theory, accounting for only the leading nonlinear terms in the wave-interaction Hamiltonian (this approach is described in the book of \cite{zakharov1992kolmogorov}).

In general, the first-order (by the value of $\delta$) nonlinear effects are presented by three-wave interaction processes: 1) when the the primary wave with wave vector $\k_1$ (hereinafter, in such cases, we write just  "the wave $\k_1$") decays into a pair of secondary waves $\k_2$ and $\k_3$ -- the decay process, 2) when two primary waves $\k_1$ and $\k_2$ merge into one secondary wave $\k_3$ -- the confluence process.

Consider, for example, the wave-decay process, which can be schematically denoted as $\k_1 \ra \k_2+\k_3$. This process is described by the conservation laws  of mechanical momentum and energy:
\begin{subequations}\label{decay law}
\begin{eqnarray}
\label{decay law1}
\k_1=\k_2+\k_3\,,
\\
\label{decay law2}
\omega_{\k_1} = \omega_{\k_2} +\omega_{\k_3}\,.
\end{eqnarray}
\end{subequations}
The fact of principal importance, which in many aspects determines the nonlinear behaviour of inertial waves (\cite{zakharov1992kolmogorov}), is that  Eqs.\,\eqref{decay law} have a nonzero set of solutions in $\k$-space, called the \emph{resonance surface}. In other words, the decay of inertial wave $\k_1 \ra \k_2+\k_3$ is \textit{allowed} by the conservation laws (in contrast to surface gravity waves, where three-wave interactions are forbidden and the first-order nonlinear effects are presented only by four-wave interactions~\citep{zakharov1992kolmogorov}). 

Inertial wave interactions can be studied by decomposition of natural variables (the velocity field) into so-called \textit{helical modes}~\citep{cambon1989spectral,waleffe1993inertial}. This approach was applied in recent years to many different problems of rotating fluids, focusing on the dynamics and statistics of inertial waves in rapidly rotating flows~(\cite{galtier2003weak,galtier2014theory}, \cite{cambon2004advances}, \cite{bellet2006wave}, see also the monograph~\cite{sagaut2008homogeneous} and references therein). However, the advantages of another well-recognised \textit{Hamiltonian approach} to study weakly nonlinear wave systems without dissipation have not been used. The key purpose of this paper is to fill this gap in the theory of rotating fluids by developing the Hamiltonian formalism for inertial waves and by discussing its advantages and future applications. This method rewrites the Euler equations as a Hamiltonian system using \textit{canonical variables} and then performs decomposition into so-called \textit{normal variables} that are analogs of the quantum-mechanical creation and annihilation operators (see the review of~\cite{zakharov1997hamiltonian} concerning the general aspects of the Hamiltonian approach). Additionally, a Hamiltonian approach was recently developed for the similar problem of \textit{internal waves} in stratified nonrotating fluids~\citep{lvov2001hamiltonian}, which provided an explanation of some features of wave energy spectra in the ocean~\citep{lvov2004energy}.

The Hamiltonian nature of the Euler equations for incompressible fluids has been known since at least the middle of the last century, including the case of rotating fluids (see the monograph of~\cite{lamb1945hydrodynamics}). More recently, this fact was discussed by~\cite{morrison1998hamiltonian} and by~\cite{salmon1988hamiltonian}. The canonical variables for an ideal fluid -- the so-called Clebsch variables -- are also well known. Therefore, the problem is mainly technical: how to introduce canonical variables for the case of rotating fluids in the most convenient way to describe inertial wave interactions.

In this paper, we first propose the Clebsch variables for rotating incompressible fluid, which provide enormous technical advantages for calculating the interaction Hamiltonian of inertial waves. We consider only inertial wave motion in the rotating reference frame when the vortex lines make no knots. In this case, according to well-known theory, the velocity field can be always represented by the Clebsch variables (see \cite{zakharov1992kolmogorov} and also the works of~\cite{yakhot1993hidden} and ~\cite{kuznetsov1980topological}).

We then focus on the wave-decay process $\k_1 \ra \k_2+\k_3$, which leads to the so-called \textit{decay instability} of inertial waves with respect to weak perturbations (see, for instance, the first paragraph of the book~\citep{Lvov1994wave}). The growth rate of the instability $\gamma(\k_1,\k_2,\k_3)$ (\textit{growth increment}) depends on the amplitude of the three-wave interactions $V^{\k_1}_{\k_2\k_3}$ and reaches its maximum on the resonance surface~\Ref{decay law}. The properties of the growth increment play a key role in the weakly nonlinear dynamics and statistics of inertial waves, which can be investigated through numerical and natural experiments (\cite{di2016quantifying}, \cite{bordes2012experimental}). In the present work, we study, in detail, the growth increment $\gamma(\k_1,\k_2,\k_3)$ on the whole three-dimensional resonance surface~\Ref{decay law} and discuss how the newly identified features can be observed experimentally. In addition, we briefly study the process of wave confluence: $\k_1+\k_2 \ra \k_3$.

Finally, we describe the turbulent cascade of inertial waves in the case of strong anisotropy, give the four-wave part of the Hamiltonian and discuss its future applications. It is important to note that the exact interaction Hamiltonian for inertial waves in rotating fluid includes only three-wave and four-wave components. Thus, we present the complete Hamiltonian formalism for inertial waves.

\section{Inertial waves of finite amplitude in rotating fluids}\label{second}

We study the Euler equations in the reference frame, rotating with an incompressible fluid:
\begin{subequations}\label{rotEuler}
\begin{eqnarray}\label{rotEuler-a}
\pt_t \v + (\v\boldsymbol{\cdot}\n)\v + 2(\bO\times\v) &=& -\n P,
\\
\label{rotEuler-b}
\n\boldsymbol{\cdot}\v &=& 0\ .
\end{eqnarray}
\end{subequations}
Here, the effective pressure $P = p - \frac12 (\bO\times\r)^2$ includes the fluid pressure $p$ and the centrifugal force $\propto \Omega^2$, while the term $2(\bO\times\v)$ is responsible for the Coriolis force~\citep{landau1987fluid}. In the main part of the paper, we work only in this rotating reference frame with the position vector $\r=(x,y,z)$ and the corresponding unit vectors $(\^\x,\,\^\y,\,\^\z)$. Without loss of generality, we choose $\bO \parallel \^\z$.

The nonlinearity of the Euler equations affects only the waves moving in different directions, which is a general consequence of incompressibility for unbounded flows (see, for instance, ~\cite{craik1986evolution}). Indeed, the plane inertial wave (as well as a package of such waves propagating in the same direction) is the exact solution of the equations~\Ref{rotEuler}. Because this fact is often overlooked in recent papers (except, to the best of our knowledge, the paper of~\cite{messio2008experimental}), we discuss it here in more detail.

We consider the general solution of Eqs.~\Ref{rotEuler} under the assumption that the velocity field $\v$ depends only on the coordinate $\xi$ along an arbitrary direction $\boldsymbol{n}$. Bearing in mind the axial symmetry around $\^\z$, we can choose $\boldsymbol{n}$ lying in the $xz$-plane so that:
\begin{equation}\label{omega}
\boldsymbol{n}=(\sin\theta,0,\cos\theta)\,, \,\,\,\,\,\,\,\,\, \xi = \r\*\boldsymbol{n}\,,  \,\,\,\,\,\,\,\,\, \v(\r,t) \equiv \v(\xi,t)=\v(x \sin\theta + z \cos\theta ,t) \,.
\end{equation}
Then, integrating the continuity equation~\Ref{rotEuler-b} by $\xi$, we find
\begin{equation}\label{v_z and v_x}
v_z = - v_x \tan\theta \,.
\end{equation}
We neglect the integration constant, which corresponds to steady motion of the whole fluid along $\^\z$. Thus, the nonlinear term in~\Ref{rotEuler-a} is exactly cancelled:
\begin{equation}
(\v\boldsymbol{\cdot}\n)\v= \biggl(v_x\sin\theta+v_z\cos\theta \biggr)\frac{\pt \v}{\pt \xi} \equiv 0\ .
\end{equation}
Now, the Eq.~\Ref{rotEuler} can be written as the following system of linear equations:
\begin{eqnarray}\label{system1}
\nonumber
\frac{\pt v_x}{\pt t} - 2\O v_y + \sin\theta \frac{\pt P}{\pt \xi} &=& 0\,,
\\
\frac{\pt v_y}{\pt t} + 2\O v_x  &=& 0\,,
\\
\nonumber
\frac{\pt v_z}{\pt t} + \cos\theta \frac{\pt P}{\pt \xi} &=& 0\,.
\end{eqnarray}
By using~\REF{v_z and v_x}, we find the exact solution for the system~\Ref{system1}:
\begin{eqnarray}\label{exact solution}
\nonumber
v_x &=& -A(\xi) \cos\theta\sin\omega t - B(\xi) \cos\theta\cos\omega t\, ,
\\
v_y &=& -A(\xi) \cos\omega t + B(\xi) \sin\omega t\, ,
\\
\nonumber
v_z &=& A(\xi) \sin\theta\sin\omega t + B(\xi) \sin\theta\cos\omega t\ .
\end{eqnarray}
Here, $\omega = 2\Omega |\cos\theta|$, while $A(\xi)$ and $B(\xi)$ are arbitrary functions. Solution~\Ref{exact solution} is the wave packet of plane inertial waves propagating along $\boldsymbol{n}$. We now introduce the wave vector $\k \parallel \boldsymbol{n}$. Choosing $A(\xi)=A\sin(\k\*\r), \,\, B(\xi)=A\cos(\k\*\r)$, we obtain the well-known circularly polarized inertial plane wave solution~\citep{landau1987fluid}:
\begin{eqnarray}\label{plane wave}
\nonumber
v_{x} &=& -A \cos\theta_{\k}\cos(\omega_{\k} t-\k\*\r),
\\
v_{y} &=& A \sin(\omega_{\k} t-\k\*\r),
\\
\nonumber
v_{z} &=& A \sin\theta_{\k}\cos(\omega_{\k} t-\k\*\r).
\end{eqnarray}
The expression~\Ref{exact solution} gives the exact class of finite amplitude solutions of the Euler equations in the rotating reference frame~\Ref{rotEuler}.

\section{Hamiltonian description of inertial waves}
\subsection{Clebsch representation of the velocity field}
The Euler equations for incompressible fluids can be written as a Hamiltonian system using the so-called Clebsch variables $\lambda(\r,t)$ and $ \mu(\r,t)$~\citep{lamb1945hydrodynamics}. Here, we find the following Clebsch representation for the velocity field $\v(\r,t)$ corresponding to the case of a rotating fluid described by~Eqs.~\Ref{rotEuler}:
\begin{eqnarray}\label{dv1}
  \v(\r,t) =\sqrt{2\Omega}\bigl[\^\y\lambda(\r,t)-\^\x\mu(\r,t)\bigr]
+\lambda(\r,t)\bn\mu(\r,t) + \n\Phi(\r,t)\,,
\end{eqnarray}
(remember that $\bO \parallel \^\z$). The potential $\Phi(\r,t)$ is uniquely determined from the appropriate boundary conditions and the continuity equation~\Ref{rotEuler-b}. Similar variables were introduced by~\cite{zakharov1971hamiltonian} and by~\cite{kuznetsov1972turbulence} in the theory of magnetized plasma. The last two terms in~\REF{dv1} correspond to the standard Clebsch representation in the case of a non-rotating fluid. The first term appears as a result of the transformation from an inertial to a rotating reference frame (see the details in the \textit{Appendix} section).

Alternatively, the velocity field $\v(\r,t)$ can be written using the transverse projector
\begin{equation}\label{projector}
\^{\P}\equiv 1-\n\Delta^{-1}\n\,,
\end{equation}
as (see the \textit{Appendix} section):
\begin{equation}\label{dv2}
  \v = \^{\P}\*\left[
    \sqrt{2\Omega}\(\lambda\^\y-\mu\^\x\) +\lambda\bn\mu
  \right]\,.
\end{equation}

The fields $\lambda(\r,t)$ and $\mu(\r,t)$ are not uniquely defined for a particular velocity field $\v(\r,t)$. For instance, the transformation with arbitrary constant parameters $a$, $b$, and $c$:
\begin{equation}\label{calibration1}
  \lambda (\r,t)\to a\lambda (\r,t)+b
\,,\qquad
  \mu (\r,t)\to \mu(\r,t) /a +c
\,,\end{equation}
does not change the field $\v$ given by~\REF{dv1}. This type of ''gauge invariance'' enables the selection of special calibration for the Clebsch variables $\lambda(\r,t)$ and $\mu(\r,t)$, which is ideal for solving a particular problem.

The fields $ \lambda(\r,t), \, \mu(\r,t)$ obey the Hamiltonian equations of motion in the canonical form:
\begin{equation}\label{eq-H}
  \p_t\lambda = \frac{\delta\H}{\delta\mu}
\,,\qquad
  \p_t\mu = -\frac{\delta\H}{\delta\lambda}
\,,\end{equation}
where $\delta$ is the variational derivative, and the Hamiltonian $\H$ is the energy represented by $\lambda$ and $\mu$:
\begin{equation}\label{Hamiltonian}
  \H = \int \frac12|\v|^2 d\r \,.
\end{equation}
Equations~\Ref{eq-H} and~\Ref{Hamiltonian} yield:
\begin{subequations}\label{eq-H1}
\begin{eqnarray}\label{eq-H1-a}
  \p_t\lambda &=& -(\v\*\bn)\lambda - \sqrt{2\Omega} v_x
\,,\\\label{eq-H1-b}
  \p_t\mu &=& -(\v\*\bn)\mu - \sqrt{2\Omega} v_y
\ .\end{eqnarray}
\end{subequations}
From Eqs.~\Ref{dv1} and~\Ref{eq-H1}, we find that $\v$ satisfies the Euler equations~\Ref{rotEuler} with the effective pressure
\begin{equation}\label{pressure}
  P = -\p_t\Phi -\lambda\p_t\mu -\frac12|\v|^2 - \frac12 (\bO\times\r)^2 \,.
\end{equation}

\subsection{Hamiltonian in a rotating reference frame}

Let us introduce the following notation for the velocity field~\Ref{dv2}:
\begin{equation}\label{v0v1}
\v = \v_1+\v_2,
\,\,\,\,\,\,\,\,\,\,\,\,
\v_1 = \^{\P}\* \[\sqrt{2\Omega}(\lambda \^\y - \mu \^\x)\],
\,\,\,\,\,\,\,\,\,\,\,\,
\v_2 = \^{\P}\*(\lambda \bn \mu),
\end{equation}
and symmetric Fourier transform:
\begin{eqnarray}
\v_{\k} = \frac{1}{(2\pi)^{3/2}}\int\v\exp(-i\k\*\r)d\r,
\,\,\,\,\,\,
\v = \frac{1}{(2\upi)^{3/2}}\int\v_{\k}\exp(i\k\*\r)d\k\,.
\end{eqnarray}
By representation~\Ref{v0v1}, we distinguish the impact of linear and nonlinear combinations ($\v_1$ and $\v_1$ respectively) of the Clebsch variables $\lambda$ and $\mu$ on the velocity field. According to~\REF{Hamiltonian} and Parseval's theorem, the Hamiltonian can be written as the sum of the following three terms:
\begin{subequations} \label{Hfull}\begin{equation}\label{H}
\H = \H_2 + \H_3 + \H_4\,.
\end{equation}
Here
\begin{eqnarray}\label{H2}
\H_2 &=& \frac{1}{2}\int |\v_{1\k}|^2d\k\
\,,\\\label{H3}
\H_3 &=& \frac{1}{2}\int (\v_{1\k}\*\v^*_{2\k}+\v^*_{1\k}\*\v_{2\k})d\k
\,,\\\label{H4}
\H_4 &=& \int |\v_{2\k}|^2 d\k
\,.
\end{eqnarray}
\end{subequations}

The interactions of inertial waves are described by $\H_3$ and $\H_4$, which we denote as \textit{interaction Hamiltonian}:
\begin{equation}\label{Hint}
\H_{\rm int} = \H_3 + \H_4\,.
\end{equation}
The expression~\Ref{Hfull} is the exact Hamiltonian for an ideal rotating incompressible fluid. It contains a finite number of the terms since for incompressible flow we use only one pair of canonical variables $\lambda(\r,t)$ and $\mu(\r,t)$ (see the corresponding discussion in~\cite{zakharov1992kolmogorov}).

Now, let us denote the components of wave vector $\k$ as:
\begin{eqnarray}\label{k spherical}
\k =(k_x,k_y,k_z) = (k\sin\theta\cos\varphi, k\sin\theta\sin\varphi, k\cos\theta),
\,\,\,\,\,\,\,\,\,\,\,\,
\k_{\bot} = (k_x,k_y).
\end{eqnarray}
Using the expression for transverse projector $\^{P}$ in $\k$-space:
\begin{equation}\label{projector_k}
\^{P}^{\alpha\beta}_{\k}=\delta^{\alpha\beta} - \frac{k^{\alpha}k^{\beta}}{k^2},
\end{equation}
and~\REF{v0v1}, we calculate the following Fourier components of the velocity field:
\begin{subequations}
\begin{eqnarray}\label{v0k}
\v_{1\k} &=& \sqrt{2\O}\bigl[\lambda_{\k} \^\y - \mu_{\k} \^\x - \frac{\k}{k^2}\bigl(k_y \lambda_{\k}-k_x\mu_{\k} \bigr) \bigr],
\\
\label{v1k}
\v_{2\k} &=& \frac{i}{(2\upi)^{3/2}}\int
\Big(\k_2-\k\frac{\k\*\k_2}{k^2} \Big)\lambda_{\k_1}\mu_{\k_2}\delta_{\k-\k_1-\k_2}d\k_1d\k_2 \,.
\end{eqnarray}
\end{subequations}
When $\k || \k_1 || \k_2$, the expression $\k_2-\k\frac{\k\*\k_2}{k^2}$ in~\REF{v1k} is exactly zero, according to the conservation law: $\k-\k_1-\k_2 = 0$. Thus, in this case, the interaction Hamiltonian $\H_{\rm int} = 0$, which again demonstrates that inertial waves moving in one direction do not interact (see the section (2)).

\subsection{Canonical form of the quadratic Hamiltonian}

We now introduce the following canonical transformation of the fields $\mu_{\k}$ and $\lambda_{\k}$ in $\k$-space:
\begin{eqnarray}\label{canonical transform}
\mu_{\k} = \wt\mu_{\k}\cos\varphi + \wt\lambda_{\k}\sin\varphi = \frac{1}{k_{\bot}} \bigl(\wt\mu_{\k} k_x + \wt\lambda_{\k} k_y \bigr)
\,,\\ \nonumber
\lambda_{\k} = -\wt\mu_{\k}\sin\varphi + \wt\lambda_k\cos\varphi = \frac{1}{k_{\bot}}\bigl(-\wt\mu_{\k} k_y + \wt\lambda_{\k} k_x \bigr)
\,.
\end{eqnarray}
Introducing the normal variables $c_{\k}$ and $c^*_{\k}$:
\begin{equation}\label{normal variables}
\wt\mu_{\k}=\sqrt{\frac{\Omega}{\omega_{\k}}}(c_{\k} + c^*_{-\k})
\,,\quad
\wt\lambda_{\k}=-\frac{i}{2}\sqrt{\frac{\omega_{\k}}{\Omega}}(c_{\k} - c^*_{-\k})\,,
\end{equation}
we present the quadratic part of the Hamiltonian $\H_2$ in diagonal form:
\begin{equation}\label{H2diagonal}
\H_2 = \int \omega_{\k}c_{\k}c^*_{\k}d\k\,,\quad \omega_{\k}=2\Omega |\cos\theta_{\k}|\ .
\end{equation}

The plane wave in normal variables corresponds to the operator
\begin{equation}\label{plane_wave_normal}
c_k = A e^{i\omega_k t}\delta(\k-\k_0)\,.
\end{equation}
By substituting~\REF{plane_wave_normal} into \REF{normal variables} and then into \REF{canonical transform} and finally calculating the linear part of the velocity field~\REF{v0k}, we recover the plane inertial wave solution~\Ref{plane wave} multiplied by scalar factor $\frac{k_x}{k_{\perp}}\sqrt{\omega_{\k}}$. This normalization of the plane wave solution in variables $c_{\k}$ and $c^*_{\k}$ should be taken into account when comparing the results of this paper with the predictions of approaches formulated with physical variables.

\subsection{Interaction Hamiltonian}\label{section interaction}
We calculate the interaction Hamiltonian $\H_{\rm int}$ in two steps. First, we calculate $\H_3$ and $\H_4$ in variables $\wt\mu_{\k}$ and $\wt\lambda_{\k}$  according to equations~(\ref{H3}), \Ref{v0k}, \Ref{v1k} and~(\ref{canonical transform}). After symmetrization, $\H_3$ takes the form:
\begin{eqnarray}\label{H3inv}
\H_3 &=& \frac{i\sqrt{2\Omega}}{2(2\upi)^{3/2}}\int\frac{d\k_1 d\k_2 d\k_3}{\kp\kpp\kppp}
\delta_{\k_1-\k_2-\k_3}
\bigl(F^{\k_1}_{\k_2\k_3}\mu_{\k_1} - S_{\k_2\k_3} \lambda_{\k_1} \bigr)\\\nonumber
&& \times\biggl[S_{\k_2\k_3} (\mu_{\k_2}\mu_{\k_3} + \lambda_{\k_2}\lambda_{\k_3}) + (\bkpp\*\bkppp) (\mu_{\k_2} \lambda_{\k_3} - \lambda_{\k_2}\mu_{\k_3}) \biggr].
\end{eqnarray}
Here, we use the auxiliary functions:
\begin{eqnarray}
F^{\k_1}_{\k_2\k_3} &=& \frac12\biggl[\kp^2 \frac{\k_1\*\k_2-\k_1\*\k_3}{k_1^2} + \kpp^2-\kppp^2 \biggr]\,,
\\\nonumber
S_{\k_2\k_3} &=& k_{3x}k_{2y} - k_{2x}k_{3y} = \kpp\kppp\sin(\varphi_2-\varphi_3)\,,
\end{eqnarray}
with the following properties:
\begin{eqnarray}\label{FSprop}
F^{\k_i}_{\k_l\k_m} = -F^{\k_i}_{\k_m\k_l},
\,\,\,\,\,\,\,\,\,
S_{\k_l\k_m} = -S_{\k_m\k_l},
\,\,\,\,\,\,\,\,\,
S_{\k_2,\k_3}=S_{\k_3,\k_1}=S_{\k_1,\k_2}.
\end{eqnarray}
For the $\H_4$ component of the interaction Hamiltonian, we find the following expression:
\begin{eqnarray}\label{H4inv}
\H_4 &=& \frac{1}{4(2\upi)^3}\int\frac{d\k_1 d\k_2 d\k_3 d\k_4}{\kp\kpp\kppp\kpppp}
\delta_{\k_1+\k_2-\k_3-\k_4} Q^{\k_1\k_2}_{\k_3\k_4}\\\nonumber
&& \times\biggl[S_{\k_1\k_2} (\mu_{\k_1}\mu_{\k_2} + \lambda_{\k_1}\lambda_{\k_2}) + (\bkp\*\bkpp) (\mu_{\k_1} \lambda_{\k_2} - \lambda_{\k_1}\mu_{\k_2}) \biggr] \\\nonumber
&& \times\biggl[S_{\k_3\k_4} (\mu_{\k_3}\mu_{\k_4} + \lambda_{\k_3}\lambda_{\k_4}) + (\bkppp\*\bkpppp) (\mu_{\k_3} \lambda_{\k_4} - \lambda_{\k_3}\mu_{\k_4}) \biggr]\,.
\end{eqnarray}
Here, we use the additional function
\begin{equation}
Q^{\k_1\k_2}_{\k_3\k_4} = \frac{(\k_1 \cdot \k_3)(\k_2 \cdot \k_4)-(\k_1 \cdot \k_4)(\k_2 \cdot \k_3)}{(\k_1+\k_2)\cdot (\k_3+\k_4)}\,.
\end{equation}
The function $Q^{\k_1\k_2}_{\k_3\k_4}$ is antisymmetric with respect to the transformations $\k_1 \leftrightarrow \k_2$  and $\k_3 \leftrightarrow \k_4$:
\begin{equation}
Q^{\k_1\k_2}_{\k_3\k_4} = -Q^{\k_2\k_1}_{\k_3\k_4}, \,\,\,\,\,\,\,\,\,\,\,\,\,\,\,\,\,\,\,\, Q^{\k_1\k_2}_{\k_3\k_4} = -Q^{\k_1\k_2}_{\k_4\k_3}.
\end{equation}
The interaction Hamiltonian is obviously invariant with respect to rotations by angle $\varphi$ (i.e., around $\^\z$), which can be directly seen from expressions~\Ref{H3inv} and~\Ref{H4inv}.

Then, we substitute the normal variables~\Ref{normal variables} into~\Ref{H3inv}, and after symmetrization, $\H_3$ takes the standard form:
\begin{eqnarray}
\H_3 &=& \frac12\int (V^{\k_1}_{\k_2\k_3} c^*_{\k_1}c_{\k_2}c_{\k_3} + c.c) \delta_{\k_1-\k_2-\k_3} d\k_1 d\k_2 d\k_3 
\\
\nonumber
&& +\frac16 \int (U_{\k_1\k_2\k_3} c^*_{\k_1}c^*_{\k_2}c^*_{\k_3} + c.c.) \delta_{\k_1+\k_2+\k_3} d\k_1 d\k_2 d\k_3 \,,
\end{eqnarray}
where
\begin{eqnarray}\label{V123_p1}
V^{\k_1}_{\k_2\k_3} &=& \frac{i \sqrt{2\Omega}}{(2\upi)^{3/2}\kp\kpp\kppp}
\biggl\{
\sqrt{\frac{\Omega^3}{\omega_{\k_1}\omega_{\k_2}\omega_{\k_3}}}  S_{\k_3\k_2} \bigl [ F^{\k_1}_{\k_2\k_3} - F^{\k_2}_{\k_1\k_3} - F^{\k_3}_{\k_2\k_1}  \bigr ]
\\
&& -\frac{3i}{8}\sqrt{\frac{\omega_{\k_1}\omega_{\k_2}\omega_{\k_3}}{\Omega^3}} S_{\k_3\k_2}^2+\frac{i}{2}\sqrt{\frac{\O}{\omega_{\k_1}\omega_{\k_2}\omega_{\k_3}}}(\omega_{\k_1}-\omega_{\k_2}-\omega_{\k_3})S_{\k_3\k_2}^2
\nonumber\\
&& +\frac{1}{4}\sqrt{\frac{\omega_{\k_2}\omega_{\k_3}}{\O \, \omega_{\k_1}}} S_{\k_3\k_2} \bigl [ -F^{\k_1}_{\k_2\k_3}  + \bkp\*\bkppp - \bkp\*\bkpp \bigr ]
\nonumber\\
&& -\frac{i}{2}\sqrt{\frac{\O \, \omega_{\k_1}}{\omega_{\k_2}\omega_{\k_3}}}\bigl [ -F^{\k_3}_{\k_2\k_1}  (\bkp\*\bkpp) + F^{\k_2}_{\k_1\k_3}  (\bkp\*\bkppp) \bigr ]
\nonumber\\
&& -\frac{i}{2}\sqrt{\frac{\O \, \omega_{\k_3}}{\omega_{\k_1}\omega_{\k_2}}}\bigl [ -F^{\k_1}_{\k_2\k_3} (\bkpp\*\bkppp) + F^{\k_2}_{\k_1\k_3}  (\bkp\*\bkppp)  \bigr ]
\nonumber\\
&& -\frac{i}{2}\sqrt{\frac{\O \, \omega_{\k_2}}{\omega_{\k_1}\omega_{\k_3}}}\bigl [ F^{\k_1}_{\k_2\k_3}  (\bkpp\*\bkppp) - F^{\k_3}_{\k_2\k_1}  (\bkp\*\bkpp)  \bigr ]
\nonumber\\
&& -\frac{1}{4}\sqrt{\frac{\omega_{\k_1}\omega_{\k_3}}{\O \, \omega_{\k_2}}} S_{\k_3\k_2} \bigl [ F^{\k_2}_{\k_1\k_3}  + \bkpp\*\bkppp + \bkp\*\bkpp  \bigr ]
\nonumber\\
&& +\frac{1}{4}\sqrt{\frac{\omega_{\k_1}\omega_{\k_2}}{\O \, \omega_{\k_3}}} S_{\k_3\k_2} \bigl [ -F^{\k_3}_{\k_2\k_1}  + \bkpp\*\bkppp + \bkp\*\bkppp  \bigr ]
\biggr\}\ .
\nonumber
\end{eqnarray}
The amplitude $V^{\k_1}_{\k_2\k_3}$ describes the three-wave interaction processes of decay and confluence, which we study in the following paragraphs. Analogous amplitudes of the three-wave interaction processes were obtained in the frame of helical mode decomposition by ~\cite{waleffe1993inertial}. However, the different structures of the variables and perturbation theory make comparison of these results a nontrivial problem that is beyond the scope of this paper.

The amplitude $U_{\k_1\k_2\k_3}$ corresponds to the so-called \textit{explosive three-wave instability}, which does not appear in the system of inertial waves. We give the exact expression for element $U_{\k_1\k_2\k_3}$ in the \textit{Appendix} for formal purposes.

The four-wave interaction Hamiltonian $\H_4$ can also be written in similar standard form after substitution of the normal variables and appropriate symmetrization:
\begin{eqnarray}\label{H4expr}
\H_4 &=& \int (W^{\k_1\k_2}_{\k_3\k_4} c^*_{\k_1}c^*_{\k_2}c_{\k_3}c_{\k_4} + c.c) \delta_{\k_1+\k_2-\k_3-\k_4} d\k_1 d\k_2 d\k_3 d\k_4 
\\
\nonumber
&& +  \int (G^{\k_2\k_3\k_4}_{\k_1} c_{\k_1}c^*_{\k_2}c^*_{\k_3}c^*_{\k_4} + c.c.) \delta_{\k_1-\k_2-\k_3-\k_4} d\k_1 d\k_2 d\k_3 d\k_4
\\
\nonumber
&& + \int R^*_{\k_1\k_2\k_3\k_4} c_{\k_1}c_{\k_2}c_{\k_3}c_{\k_4} \delta_{\k_1+\k_2+\k_3+\k_4} d\k_1 d\k_2 d\k_3 d\k_4\,.
\end{eqnarray}
The coefficients $W^{\k_1\k_2}_{\k_3\k_4}$, $G^{\k_2\k_3\k_4}_{\k_1}$ and $R^*_{\k_1\k_2\k_3\k_4}$, which correspond to $\k_1 + \k_2 \ra \k_3 + \k_4$, $\k_1 + \k_2 + \k_3 \ra \k_4$ and $\k_1 + \k_2 + \k_3 + \k_4 \ra 0$ four-wave  interaction processes, are given in the \textit{Appendix} section. The final representation of the Hamiltonian obviously does not depend on the chosen calibration of the Clebsch variables (see \REF{calibration1} and \REF{calibration2}).

\section{Resonance surface}
Wave decay $\k_1 \ra \k_2 + \k_3$ is the key weakly nonlinear process that determines the dynamics of inertial waves. For a given wave vector $\k_1$, the momentum and the energy conservation laws~(\ref{decay law1},\ref{decay law2}) define a two-dimensional resonance surface in $\k$-space (we have only two free parameters). Assuming that $\k_1$ lies in the $xz$-plane and using the momentum conservation law~(\ref{decay law1}) to eliminate the unknown components of the vector $\k_2$, we find the following parametrization for vectors $\k_1,\k_2,\k_3$:
\begin{eqnarray}\label{decay k resonance}
\k_1 &=& (k_1\sin\theta_1 \,\,, 0 \,\,, k_1\cos\theta_1),
\\ \nonumber
\k_2 &=& (k_1\sin\theta_1 - k_{3x} \,\,,- k_{3y} \,\,, k_1\cos\theta_1 - k_{3z})\,,
\\ \nonumber
\k_3 &=& (k_{3x} \,\,, k_{3y} \,\,, k_{3z})\ .
\end{eqnarray}
Then, using the energy conservation law~\Ref{decay law2} and the dispersion relation~\Ref{dispersion}, we obtain the following resonance condition:
\begin{eqnarray}\label{resonance surface-eq}
|k_1\cos\theta_1 |=\frac{|k_{3z}|}{\sqrt{k_{3x}^2+k_{3y}^2+k_{3z}^2}}
+\frac{|k_1cos\theta_1-k_{3z}|}{\sqrt{(k_1\sin\theta_1 - k_{3x})^2+k_{3y}^2+(k_1\cos\theta_1 - k_{3z})^2}}\ .
\end{eqnarray}
The relation~\Ref{resonance surface-eq} defines a two-dimentional set of vectors $\k_3$, which gives the main contribution to the decay processes (obviously the same resonance surface is also valid for the vector $\k_2$). The resonance surface~\Ref{resonance surface-eq} was studied by~\cite{bellet2006wave}, and the experimental verification of the resonance condition~\Ref{decay k resonance} in a rotating water tank was recently demonstrated by~\cite{bordes2012experimental}.

\begin{figure}
\includegraphics[width=2.5in]{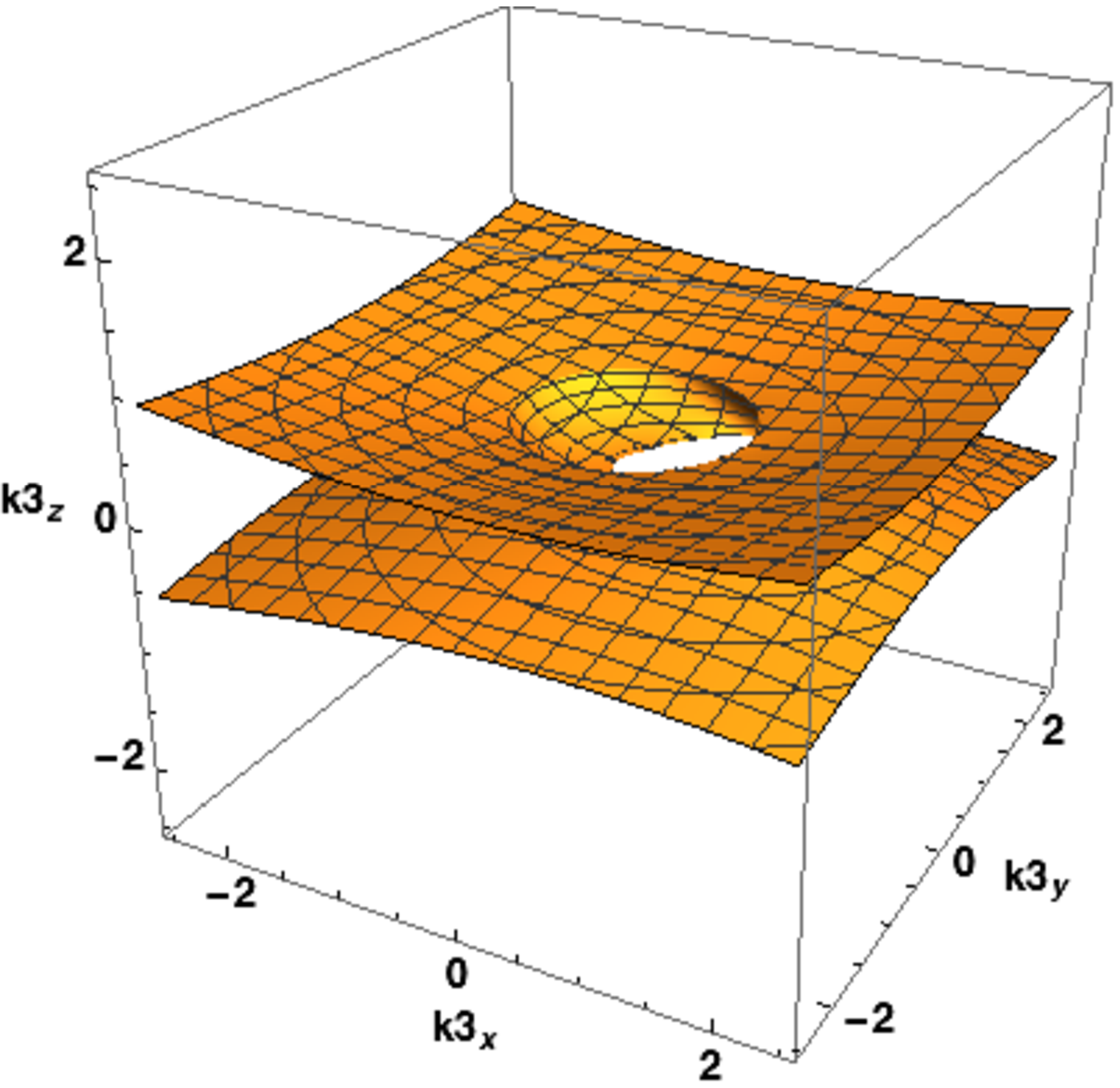}
\includegraphics[width=2.5in]{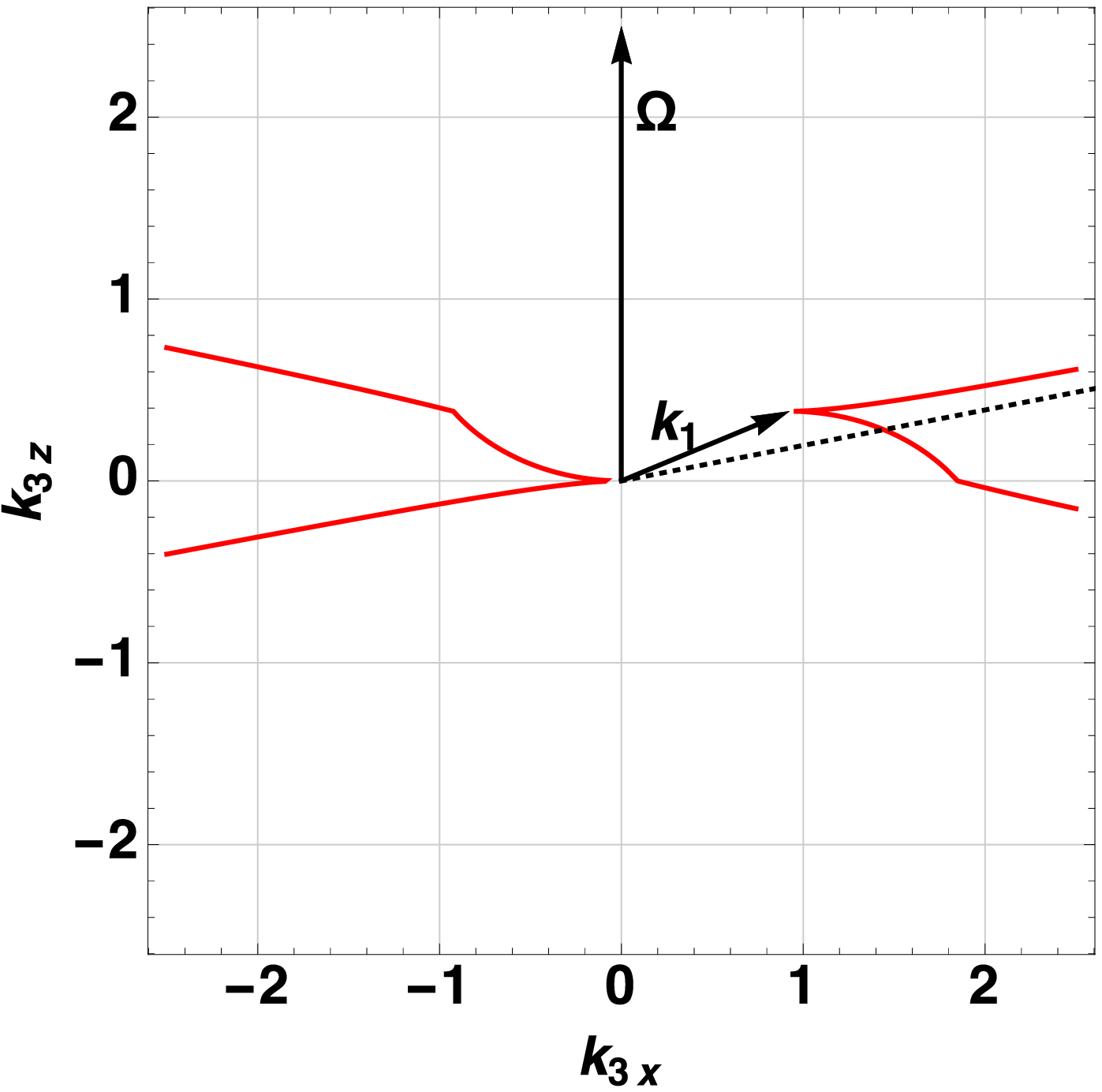}
\caption{\label{resonance_surface}
LEFT: The resonance surface of the wave vector $\k_3$ for the decay process $\k_1 \ra \k_2+\k_3$. The primary wave vector $\k_1=\bigl(\sin\frac{3\pi}{8},0,\cos\frac{3\pi}{8} \bigr)$. RIGHT: The cut of this surface (resonance curve) along the $k_xk_z$-plane.}
\end{figure}

Since the expression~\Ref{resonance surface-eq} cannot be resolved explicitly, we find solutions numerically using "Wolfram Mathematica". In Fig.~\ref{resonance_surface} we present an example of a resonance surface and its two-dimensional cut in a specific direction -- the so-called \textit{resonance curve}. The energy of the primary wave $\k_1$ is always larger than the energy of the secondary waves $\k_2$ and $\k_3$. Thus, according to the dispersion relation~\Ref{dispersion}, the directions of the secondary waves are always closer to the $xy$-plane than the direction of the vector $\k_1$ -- see the Fig.~\ref{resonance_surface}~(right).

\begin{figure}\center
\includegraphics[width=2.2in]{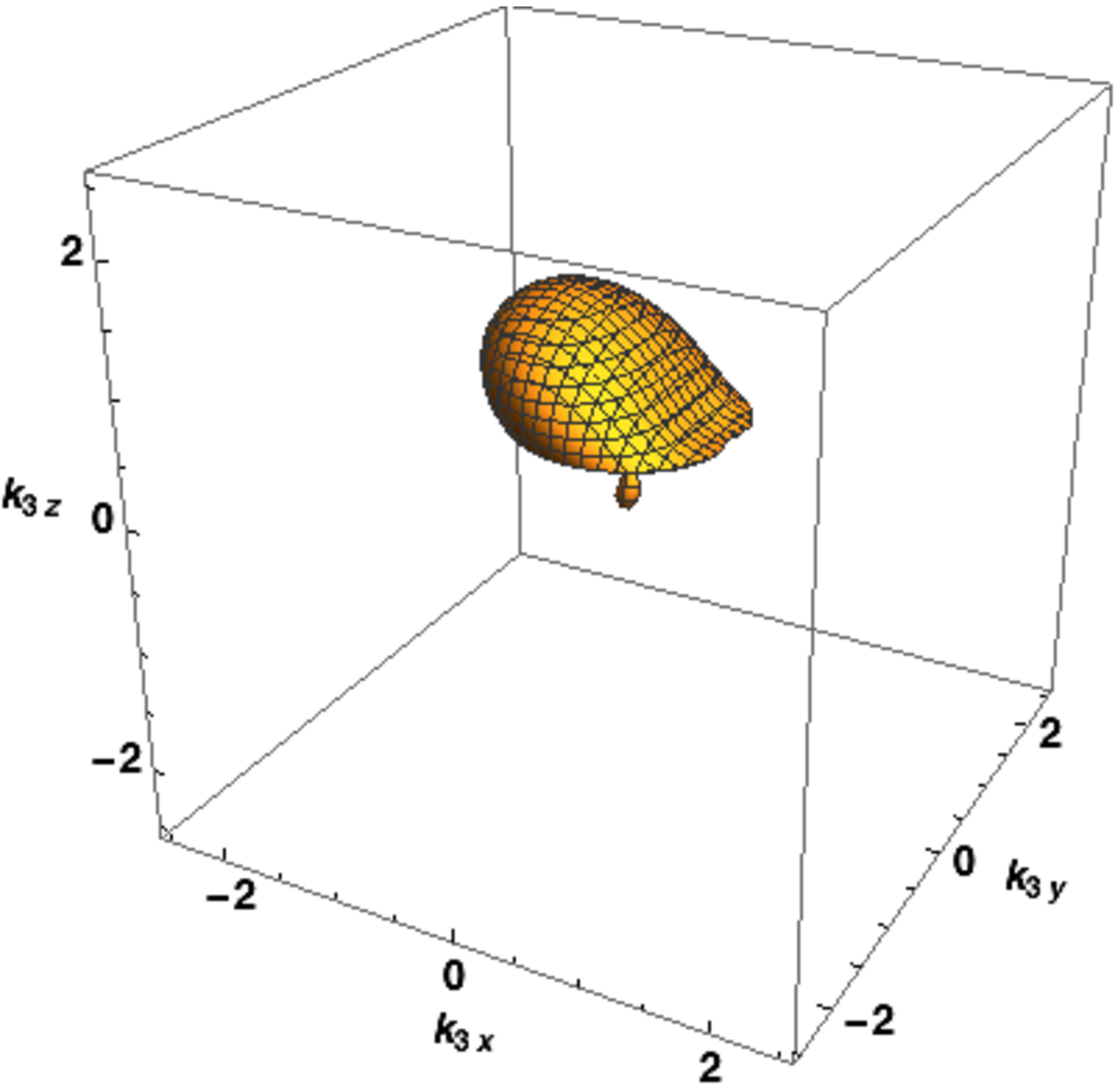}
\includegraphics[width=2.2in]{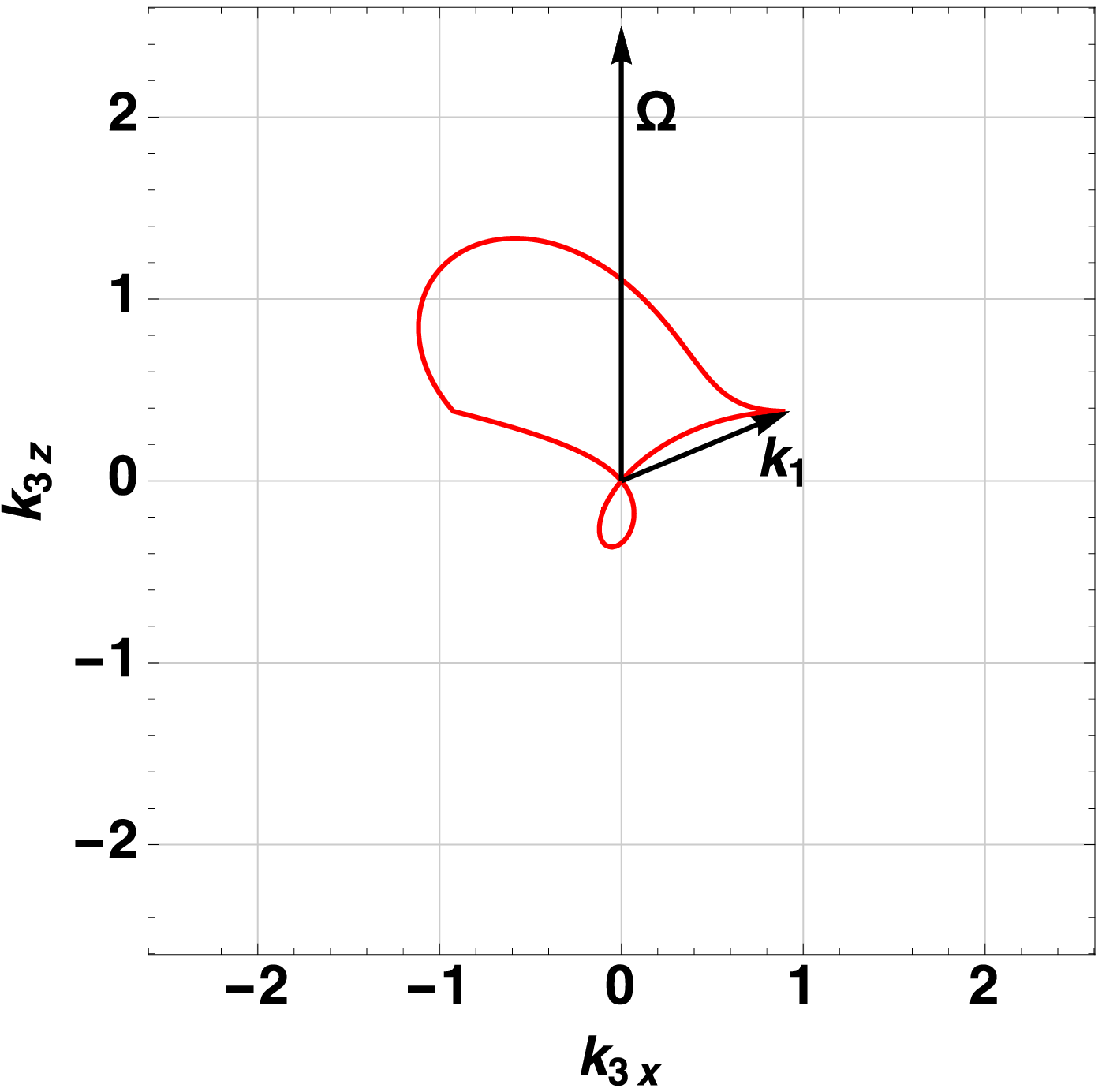}
\caption{\label{confluence_resonance_surface}
LEFT: The resonance surface of wave vector $\k_3$ for the confluence process $\k_1 + \k_2 \ra \k_3$. The primary wave vector 
$\k_1=\bigl(\sin\frac{3\pi}{8},0,\cos\frac{3\pi}{8} \bigr)$. RIGHT: The cut of this surface (resonance curve) along the $k_xk_z$-plane.}
\end{figure}

The resonance surface, having a nontrivial form with branches at $k \ra \infty$, formally allows decay into secondary waves of arbitrary small wavelength (in the experiment, this is obviously limited by the dissipation rate). We find the following asymptotics for the infinite branches assuming~\Ref{resonance surface-eq} $k_3 \ra \infty$:
\begin{equation}\label{asymptotic}
\frac{|k_1\cos\theta_1|}{2}=\frac{|k_{3z}|}{\sqrt{k_{3x}^2+k_{3y}^2+k_{3z}^2}} \,,
\end{equation}
(see Fig.~\ref{resonance_surface}~(right)). In addition, we study the process of wave confluence
\begin{equation}\label{confluence}
\k_1 + \k_2 \ra \k_3\,.
\end{equation}
In Fig.~\ref{confluence_resonance_surface}, we plot the corresponding resonance surface that is obtained similarly to the~\REF{resonance surface-eq} relation for the process~\Ref{confluence}. The confluence resonance surface remains finite in $\k$-space and grows in size when $\theta_1$ tends to $\pi/2$.

\section{Decay instability}
The decay process $\k_1 \ra \k_2+\k_3$, which is allowed by the dispersion law~\Ref{dispersion}, leads to instability of the inertial waves with respect to weak perturbations. In the general weakly nonlinear wave system with the decay-type dispersion law, the perturbations grow exponentially in the initial (linear) stage of this \textit{decay instability} (see, e.g.,~\cite{zakharov1992kolmogorov}). Consider at $t=0$ the primary wave of amplitude $A_{\k_1}$ and secondary waves of small amplitudes $A_{\k_2}$ and $A_{\k_3}$ playing the role of perturbations: $A_{\k_1}(0) \gg \max(|A_{\k_2}(0)|,\, |A_{\k_3}(0)|)$. Then, according to the well-known theory (see, e.g., Sec.1 in~\cite{Lvov1994wave}), the amplitudes of the secondary waves grow with time $t$ as:
\begin{equation}
A_{\k_{2,3}}(t)=A_{\k_{2,3}}(0)\exp(\lambda t)\,, 
\end{equation}
during the linear stage of the decay instability (meanwhile $A_{\k_1}\, \sim $\, const). Here,
\begin{equation}
\lambda = \frac{i\Delta \omega_{\k }}{2} + \gamma\,, \quad \Delta \omega_{\k } \equiv \omega_{\k_1}-\omega_{\k_2}-\omega_{\k_3}\,.
\end{equation}
The \textit{growth increment} of the decay instability
\begin{equation}\label{growth_increment}
\gamma = \sqrt{ 4|V^{\k_1}_{\k_2\k_3}|^2 |A_{\k_1}|^2 - \frac{\Delta \omega_{\k }^2}{4}}\,,
\end{equation}
is localized in the layer (in $\k$-space) of thickness $8|V^{\k_1}_{\k_2\k_3}|^2 |A_k(0)|^2$ near the resonance surface $\Delta \omega_{\k }=0$~\Ref{resonance surface-eq}, where it reaches the  maximum value:
\begin{equation}\label{increment_max}
\gamma_{\rm max}=2|V^{\k_1}_{\k_2\k_3}||A_k(0)|\,.
\end{equation}

\begin{figure}
\includegraphics[width=2.5in]{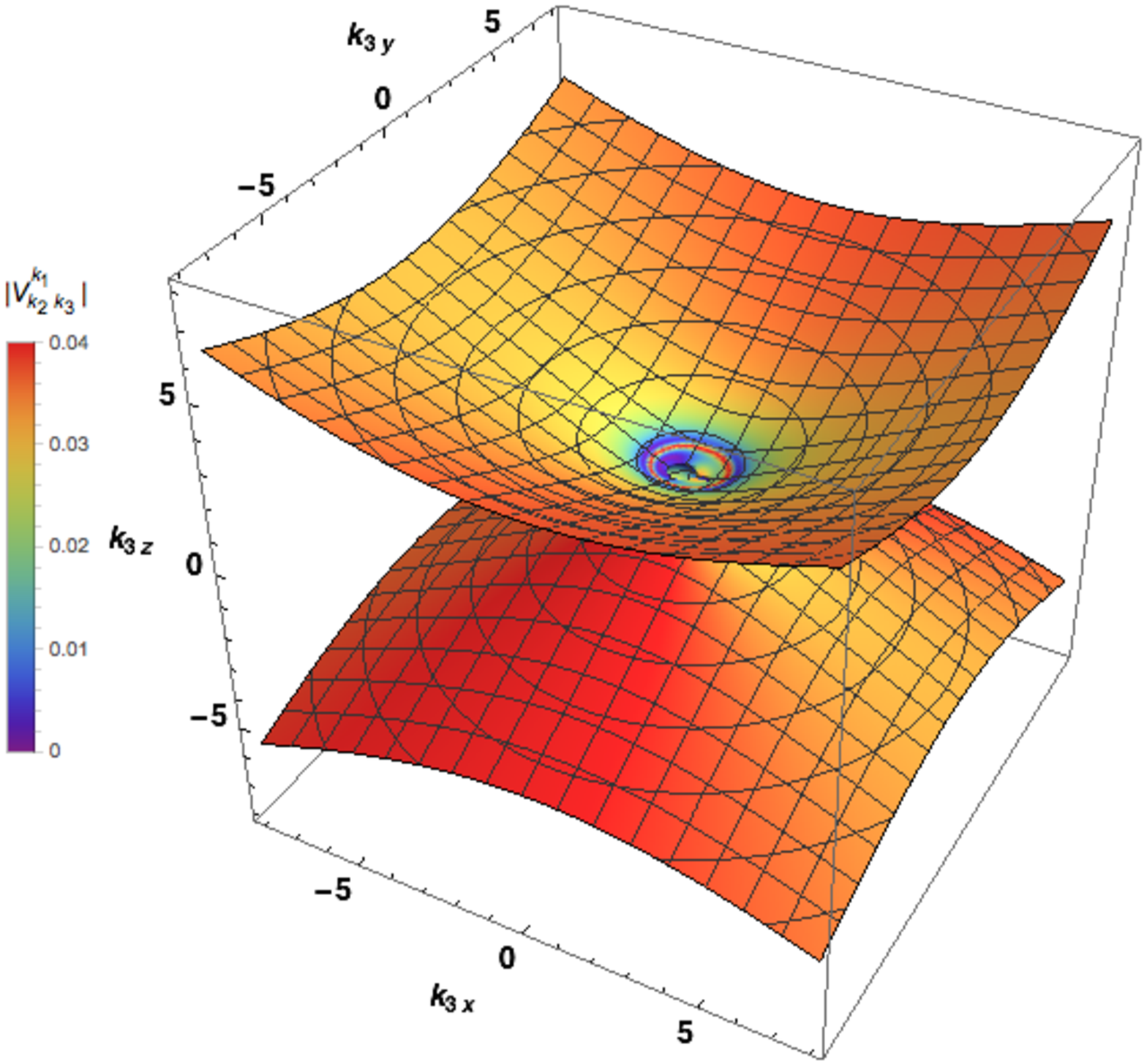}
\includegraphics[width=2.5in]{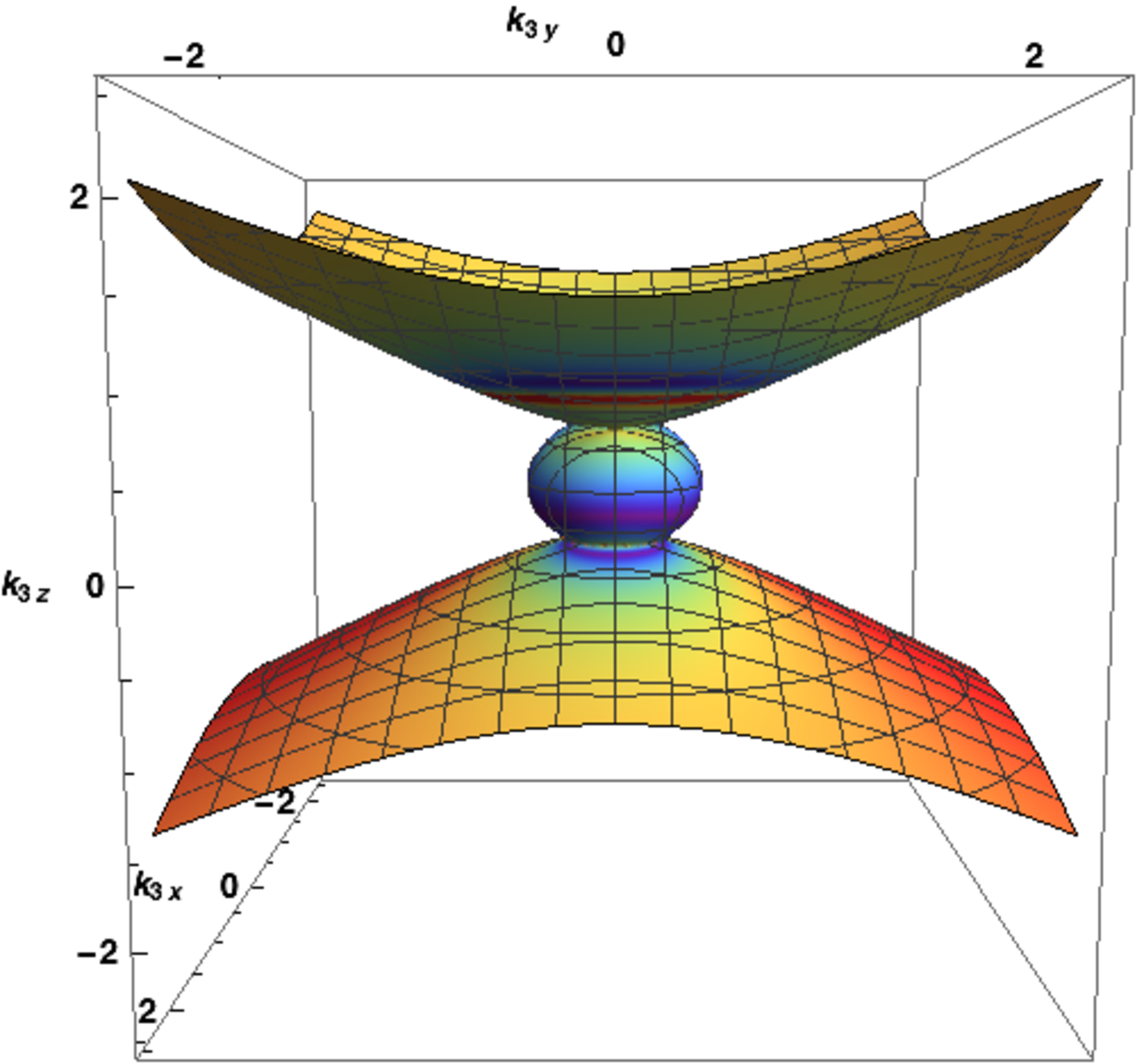}
\\
\includegraphics[width=2.5in]{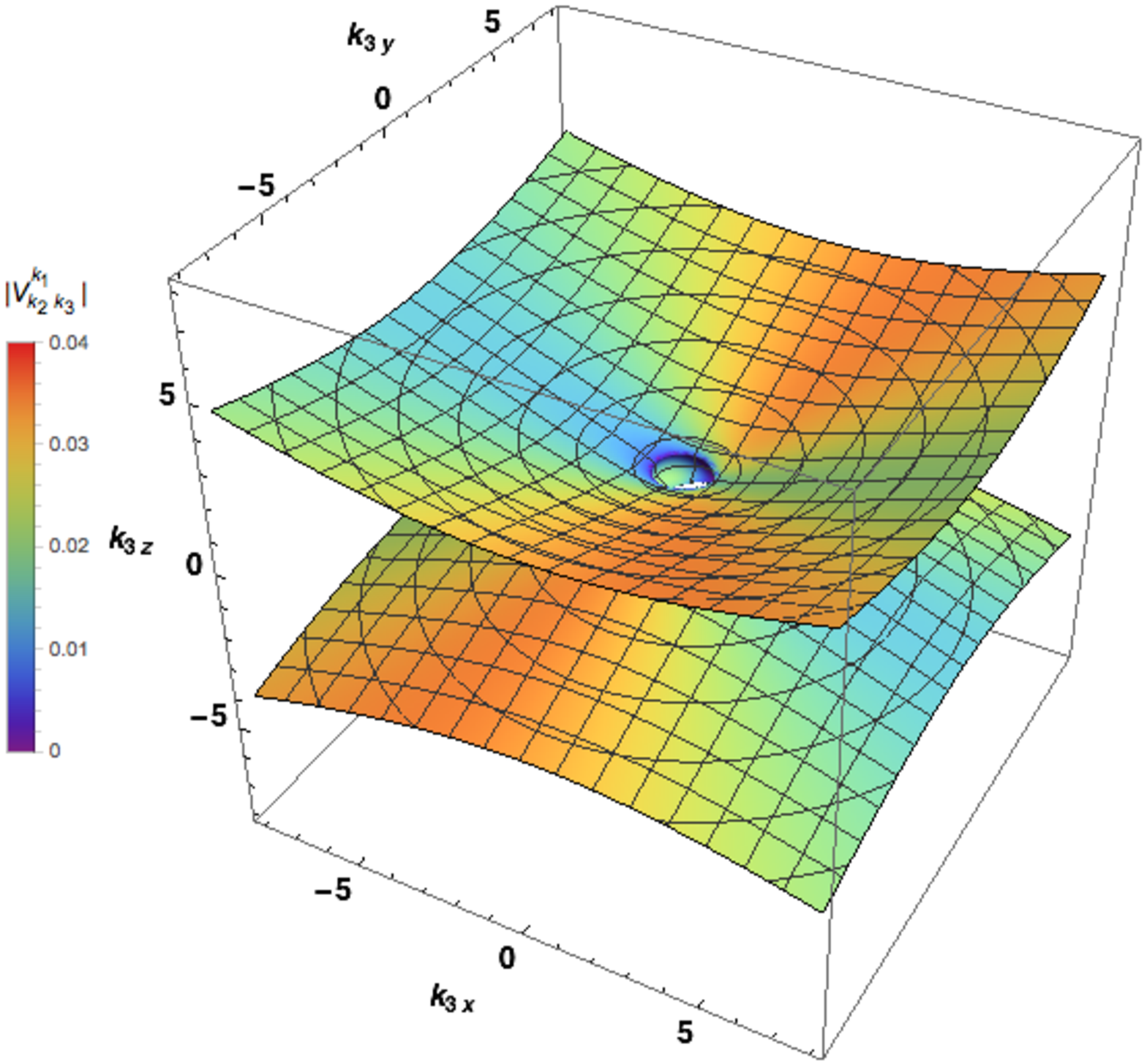}
\includegraphics[width=2.5in]{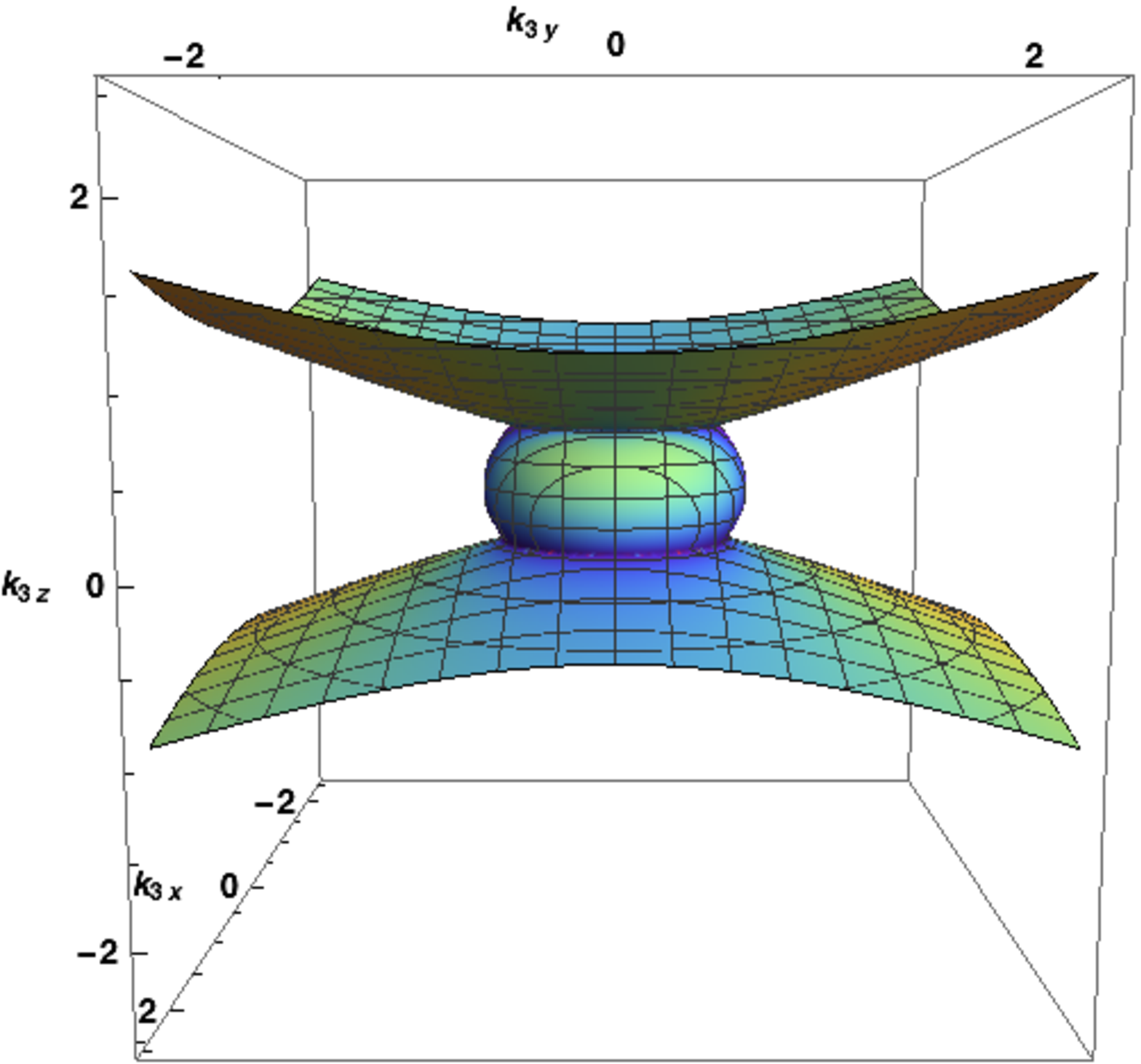}
\\
\includegraphics[width=2.5in]{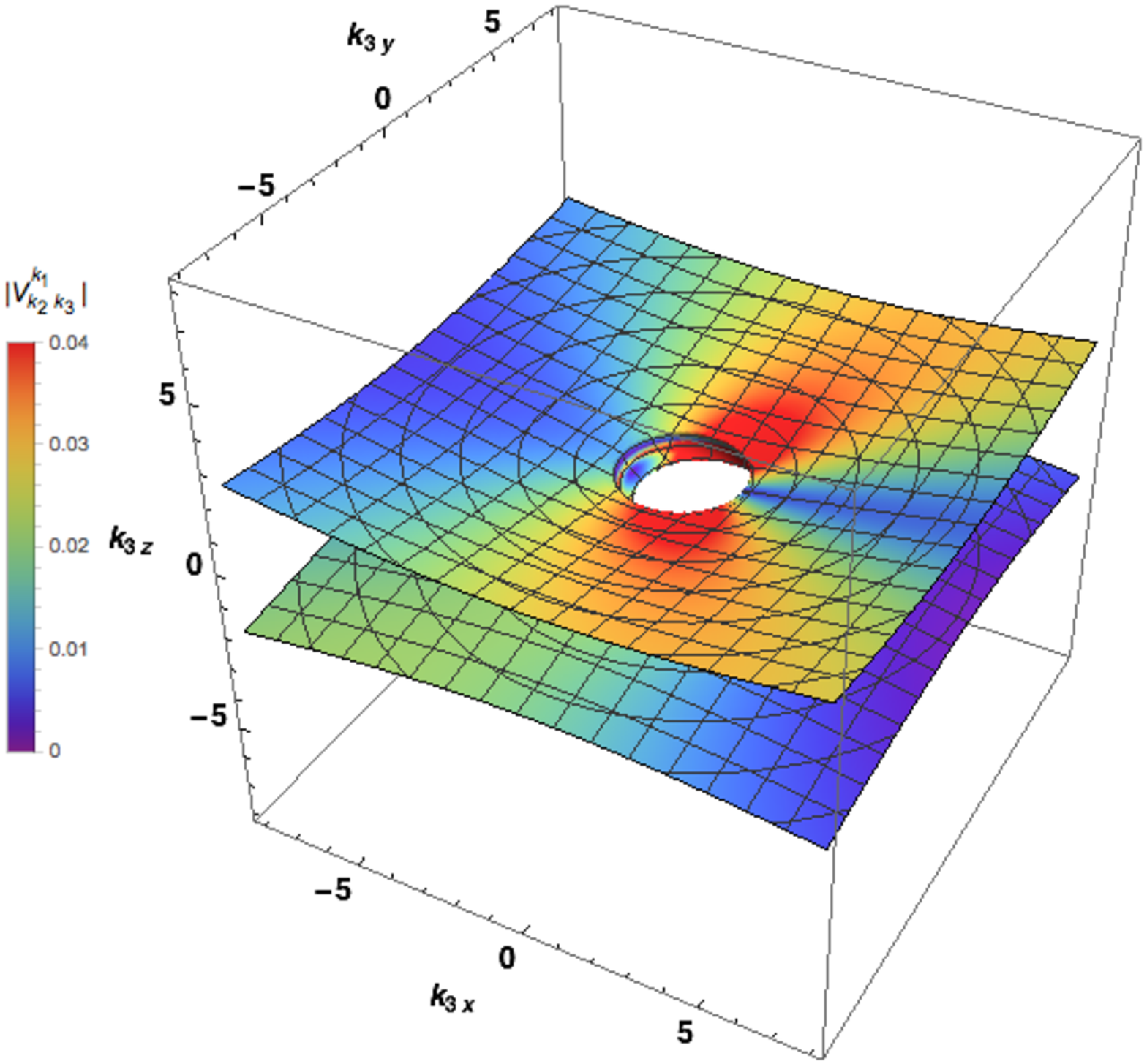}
\includegraphics[width=2.5in]{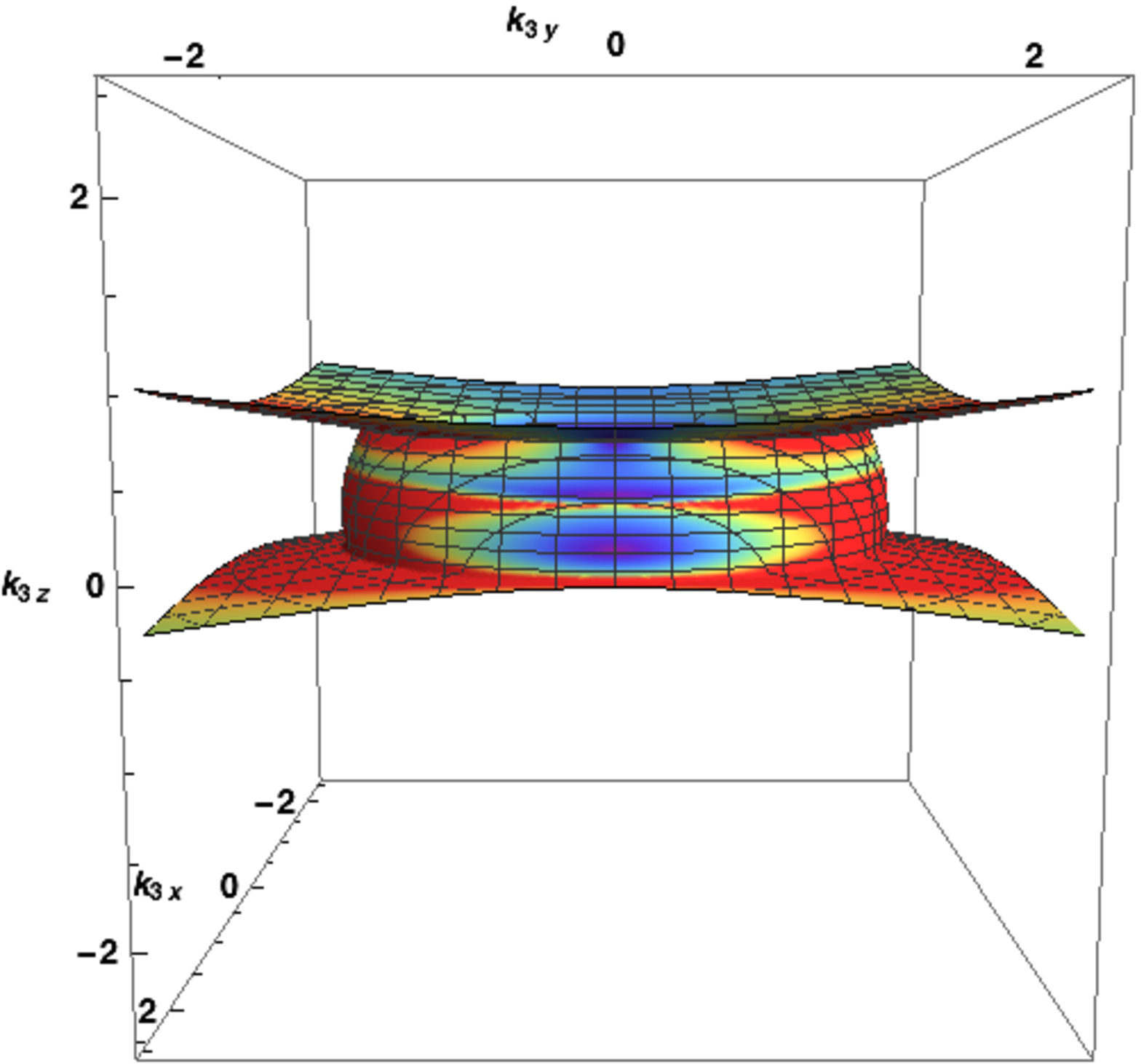}
\caption{\label{color_resonance_surface}
The resonance surface of the wave vector $\k_3$ for the decay process $\k_1 \ra \k_2+\k_3$, colored according to the value of $|V^{\k_1}_{\k_2\k_3}|$. Top row: $\k_1=\bigl(\sin\frac{3\pi}{8},0,\cos\frac{3\pi}{8} \bigr)$, middle row: $\k_1=\bigl(\sin\frac{\pi}{4},0,\cos\frac{\pi}{4} \bigr)$, bottom row: $\k_1=\bigl(\sin\frac{\pi}{8},0,\cos\frac{\pi}{8} \bigr)$. LEFT and RIGHT images show the same surface at different scales and from different view points.}
\end{figure}

In this section, we study the absolute value of the three-wave decay amplitude $|V^{\k_1}_{\k_2\k_3}|$~\Ref{V123_p1} at the resonance surface~\Ref{resonance surface-eq}, which is sufficient to describe the main features of the instability increment behaviour. In Fig.~\ref{color_resonance_surface}, we present the typical examples of a resonance surface, colored according to the value of $|V^{\k_1}_{\k_2\k_3}|$.

\begin{figure}
\center
\includegraphics[width=2in]{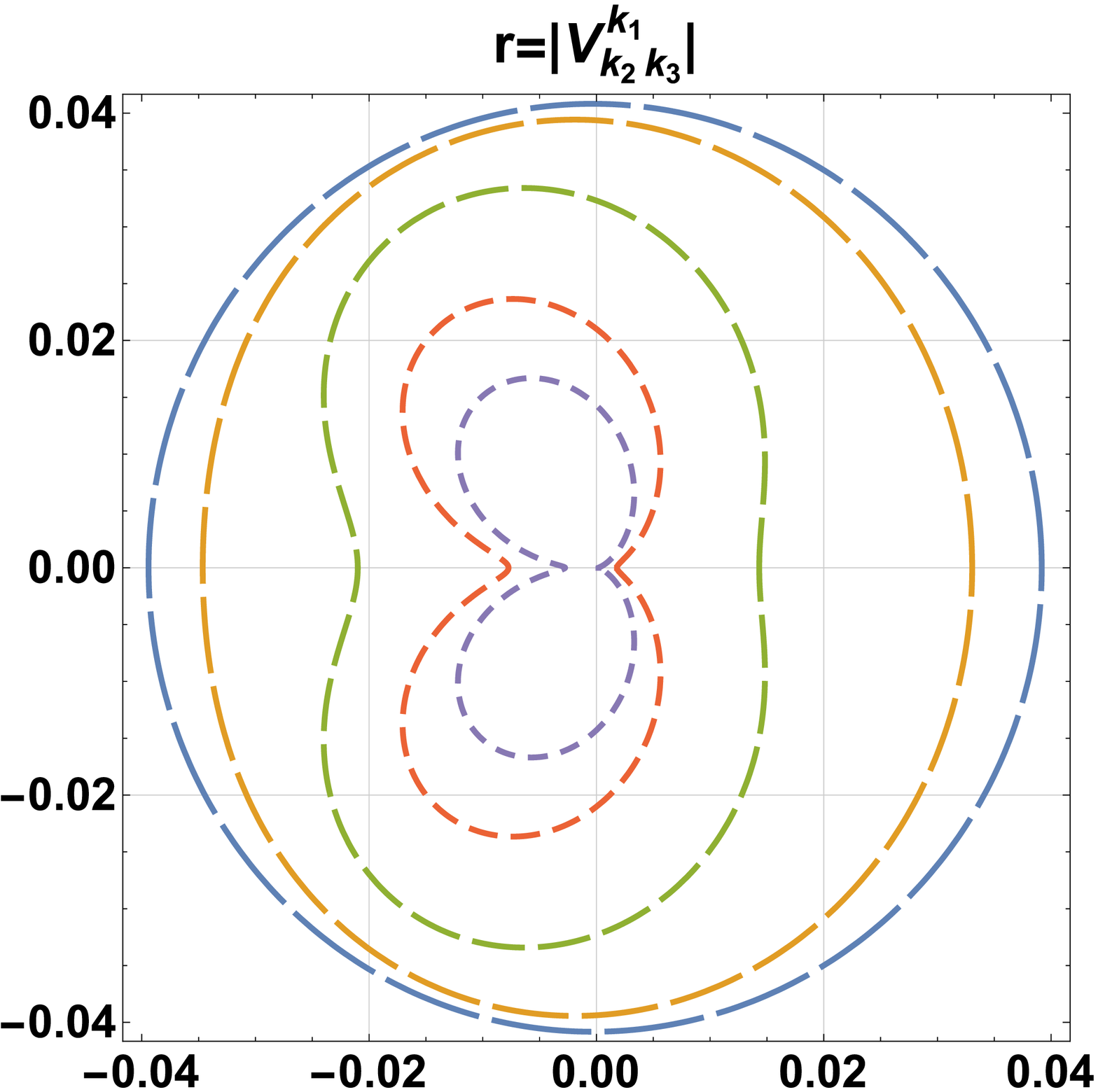}
\includegraphics[width=2.65in]{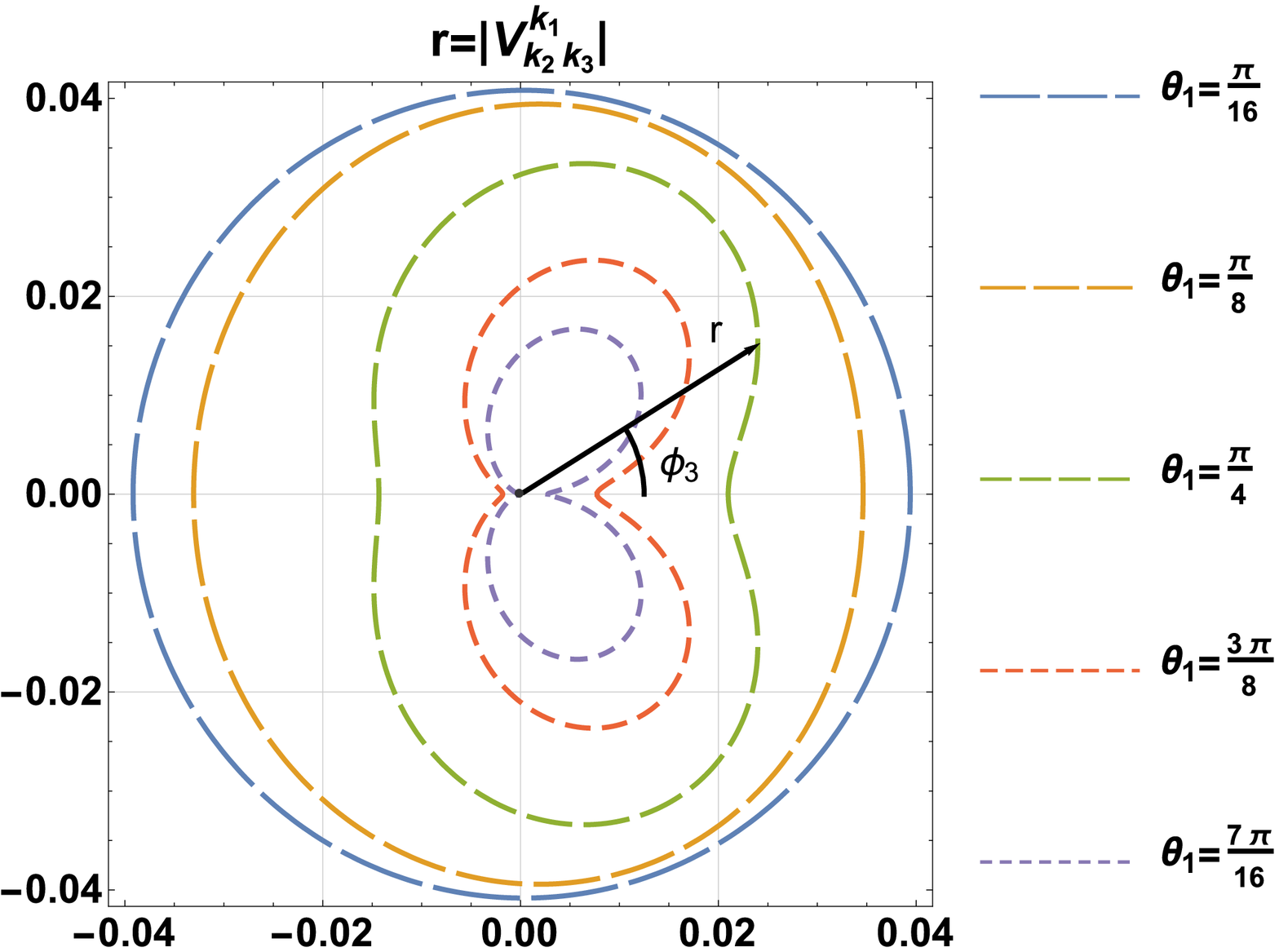}
\caption{\label{asymploticfromangle}
The polar diagram for the shortwave asymptotics \Ref{asimptotic_full} of the amplitude $|V^{\k_1}_{\k_2\k_3}(\varphi_3)|$ on the resonance surface. The primary wave vector $\k_1=\bigl(\sin\theta_1,0,\cos\theta_1 \bigr)$. The amplitude at a certain angle $\varphi_3$ is given by the length of the corresponding radius vector. The left side corresponds to the lower branch of the resonance surface; the right side corresponds to the upper branch.}
\end{figure}

We find that the decay amplitude has a nonzero value on the infinite branches of the resonance surface (see the Fig.~\ref{color_resonance_surface}). Thus, decay instability appears even for shortwave perturbations up to the dissipative threshold. We calculate the shortwave asymptotic expression of the decay amplitude $V^{\k_1}_{\k_2\k_3}$ on the upper infinite branch of the resonance surface using equations~(\ref{V123_p1}, \ref{decay k resonance}, \ref{resonance surface-eq}, \ref{asymptotic}):
\begin{eqnarray}\nonumber 
&& V^{\k_1}_{\k_2\k_3}(\varphi_3) \xrightarrow[\omega_{\k_1}=\omega_{\k_2}+\omega_{\k_3}]{k_3 \ra \infty} \frac{i \sqrt{2\Omega}}{(2\pi)^{3/2}}
\biggl\{
-\frac{1}{4}\sqrt{\frac{\omega_{\k_1}}{\Omega}}\sin\varphi_3\cos\varphi_3\sin\theta_1 \biggl(1+\frac{5}{16}\frac{\omega_{\k_1}^2}{\Omega^2} \biggr)
\\ \label{asimptotic_full}
&&~~ -\frac12\sqrt{\frac{\Omega}{\omega_{\k_1}}}\frac{\sin\varphi_3\cos\theta_1}
{\sqrt{1- \omega_{\k_1}^2/16\Omega^2}}
\biggl (1-\frac{11}{16}\frac{\omega_{\k_1}^2}{\Omega^2}+\frac{5}{128}\frac{\omega_{\k_1}^4}{\Omega^4}  \biggr )
\\\nonumber
&&~~ -\frac{3i}{16} \biggl(\frac{\omega_{\k_1}}{\Omega} \biggr)^{3/2}\sqrt{1-\frac{\omega_{\k_1}^2}{4\Omega^2}}\sin^2\varphi_3
+\frac{i}{2}\biggl( \frac{\omega_{\k_1}}{4\Omega}\sin\theta_1\cos\varphi_3-\sqrt{1-\frac{\omega_{\k_1}^2}{16\Omega^2}}\cos\theta_1 \biggl)
\\\nonumber
&&~~ \times\biggl [\frac{3}{2}\sqrt{\frac{\omega_{\k_1}}{\Omega}}\cos\varphi_3 - \sqrt{\frac{\Omega}{\omega_{\k_1}}\biggl(1-\frac{\omega_{\k_1}^2}{16\Omega^2}\biggr)}
\biggl (\frac{\omega_{\k_1}}{\Omega}\sqrt{1-\frac{\omega_{\k_1}^2}{16\Omega^2}}\cos\varphi_3 - \cos\theta_1\sin\theta_1 \biggr) \biggr ]
\biggr\}\ .
\end{eqnarray}
The shortwave asymptotics for the lower infinite branch of the resonance surface can be obtained from the expression~\Ref{asimptotic_full} by the transformation $\varphi_3 \ra \varphi_3+\pi$.
In Fig.~\ref{asymploticfromangle}, we present the angular dependence of the asymptotic~\Ref{asimptotic_full} for different values of the primary vector $\k_1$. The shortwave asymptotic is almost isotropic around the rotational axes $\bO \| \^\z$ when the direction of $\k_1$ is close to the $xy$-plane (i.e., the energy of the primary wave is low). In the opposite limit -- when the primary wave propagates almost parallel to the rotational axes -- the shortwave asymptotic is anisotropic with maxima near $\varphi_3 = \pi/4; \, 3\pi/4$, but its characteristic amplitude decreases.

\begin{figure}
\includegraphics[width=2.6in]{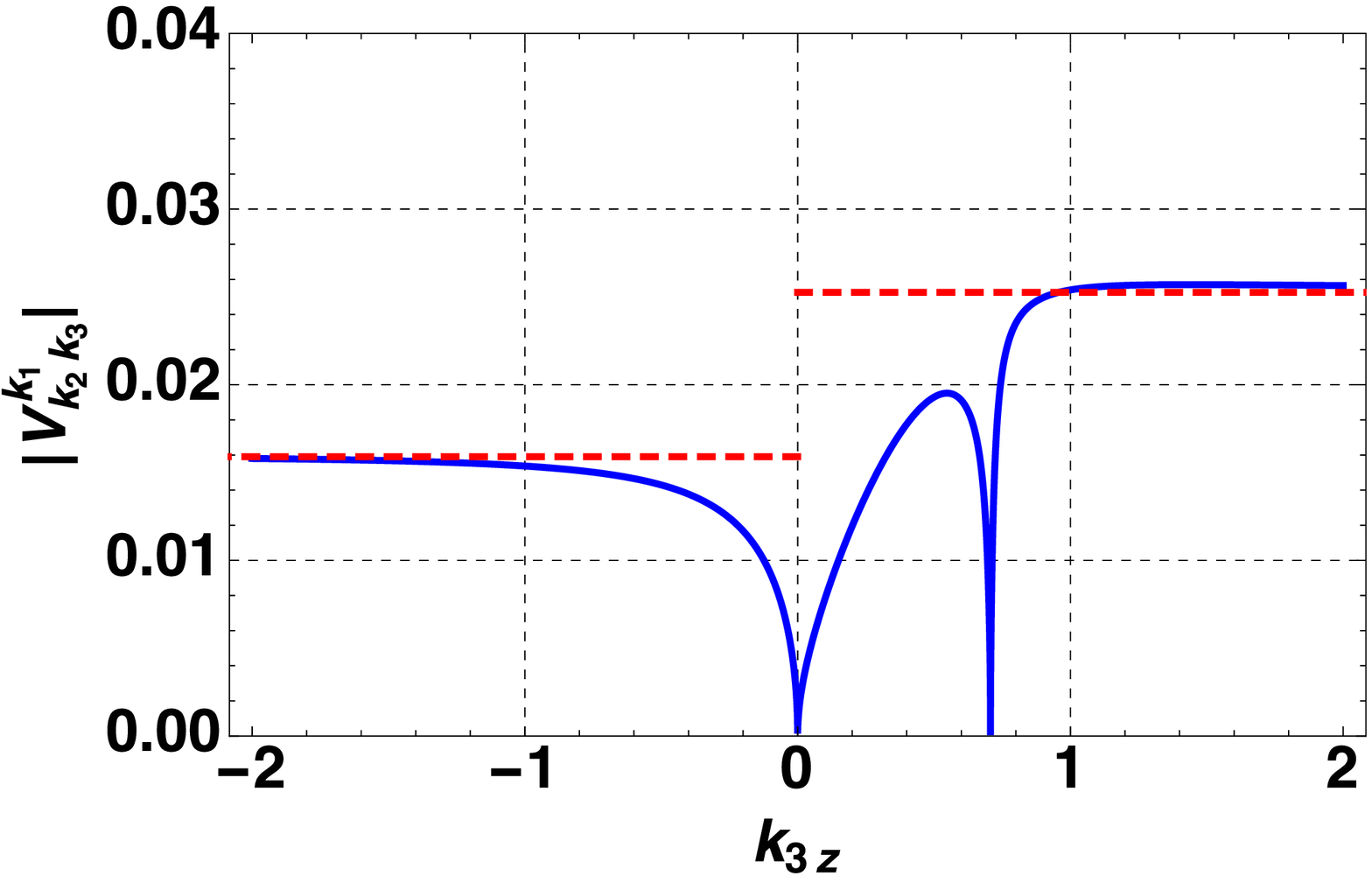}
\includegraphics[width=2.6in]{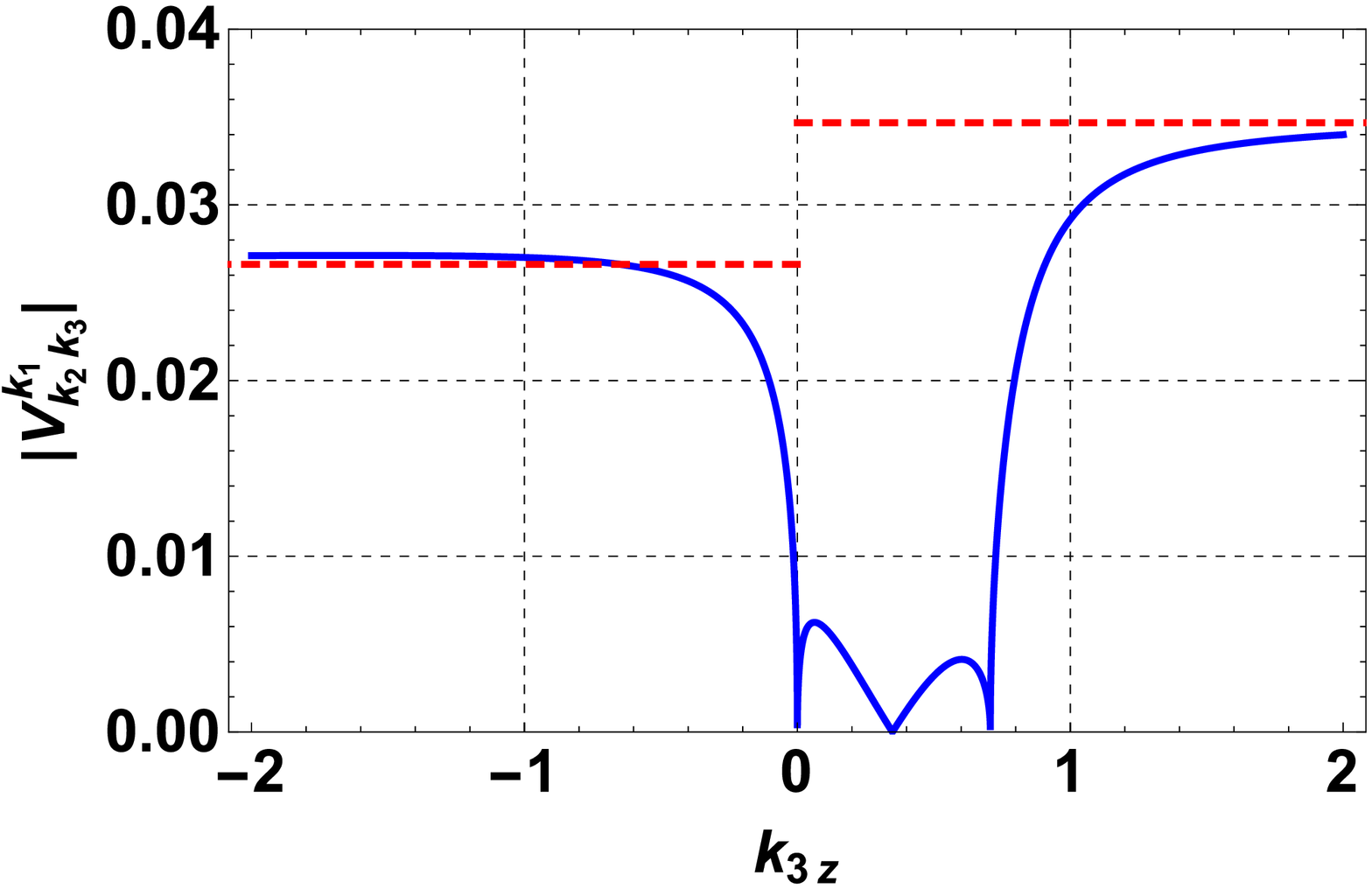}
\caption{\label{1Dfirst}
The behaviour of the amplitude $|V^{\k_1}_{\k_2\k_3}|$ (blue solid lines) on the resonance curve as a function of $k_{3z}$.  The primary wave vector 
$\k_1=\bigl(\sin\frac{\pi}{4},0,\cos\frac{\pi}{4}\bigr)$. LEFT: the resonance curve at $\varphi_1=\frac{\pi}{8}$. The amplitude has a local maximum at the middle branch of the resonance curve and a minimum at the point where one branch of the resonance curve changes to another; then, it approaches to the shortwave asymptotic~\Ref{asimptotic_full} (red dotted lines). RIGHT: the resonance curve at $\varphi_1=\frac{3\pi}{8}$. The growth increment has two local maxima and one minimum in the middle branch of the resonance curve. Again, the shortwave asymptotic~\Ref{asimptotic_full} is represented by the red dotted curve.}
\end{figure}

Fig.~\ref{resonance_surface} demonstrates that the right ($k_{3x}>0$) and left ($k_{3x}<0$) parts of the decay resonance curve can be represented as a single-valued function of $k_{3z}$. This enables us to plot the values of the amplitude~\Ref{V123_p1} on the resonance curve as a function of one variable -- see Fig.~\ref{1Dfirst}.

As we will discuss in the next section, the anisotropic case, when $k_z \ll k_{\perp}$, plays a central role in the statistical behaviour of inertial waves. Such waves, propagating predominantly in the $\k_{\perp}$-direction, have low energy and  frequency, according to the dispersion law~\Ref{dispersion}. Here, we find the following asymptotics for the amplitude $V^{\k_1}_{\k_2\k_3}$ in this \textit{small frequency limit}:

\begin{eqnarray}\label{V123 asymptotic small omega}
V^{\k_1}_{\k_2\k_3} \approx
\frac{i}{16 \upi^{3/2}}\frac{S_{\k_2\k_3}}{\kp\kpp\kppp \sqrt{\omega_{\k_1}\omega_{\k_2}\omega_{\k_3}}}
\\\nonumber
\times\biggl [\omega_{\k_1}^2 (\kppp^2-\kpp^2)+\omega_{\k_2}^2 (\kp^2-\kppp^2)-\omega_{\k_3}^2 (\kp^2-\kpp^2) \biggr ]\,, \,\,\,\,\,\,\,\,\,\,\,\,\,\,\,\,\,\, \text{at   }k_{z} \ll k_{\perp} \,.
\end{eqnarray}

\begin{figure}
\includegraphics[width=2.5in]{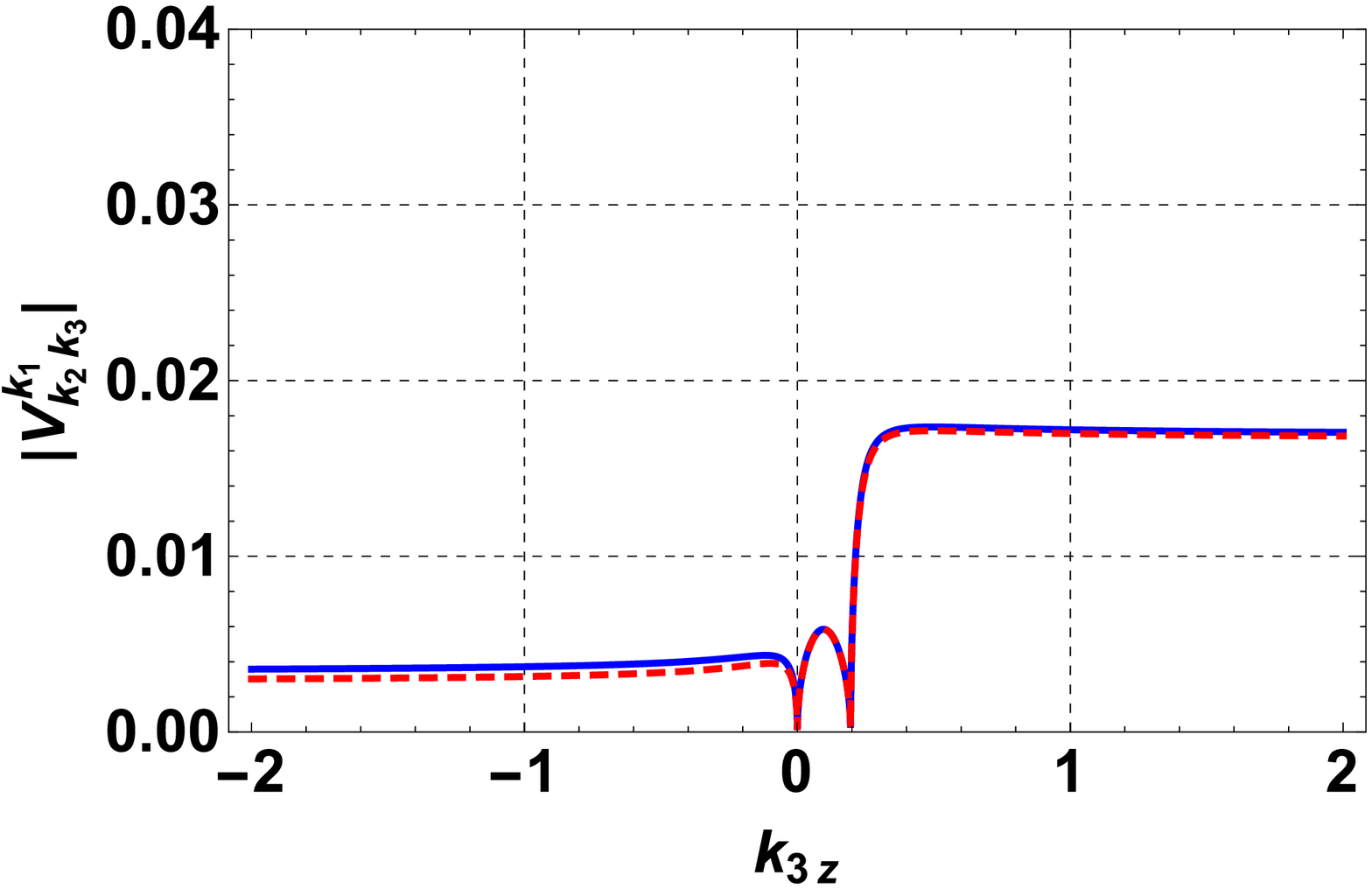}
\includegraphics[width=2.5in]{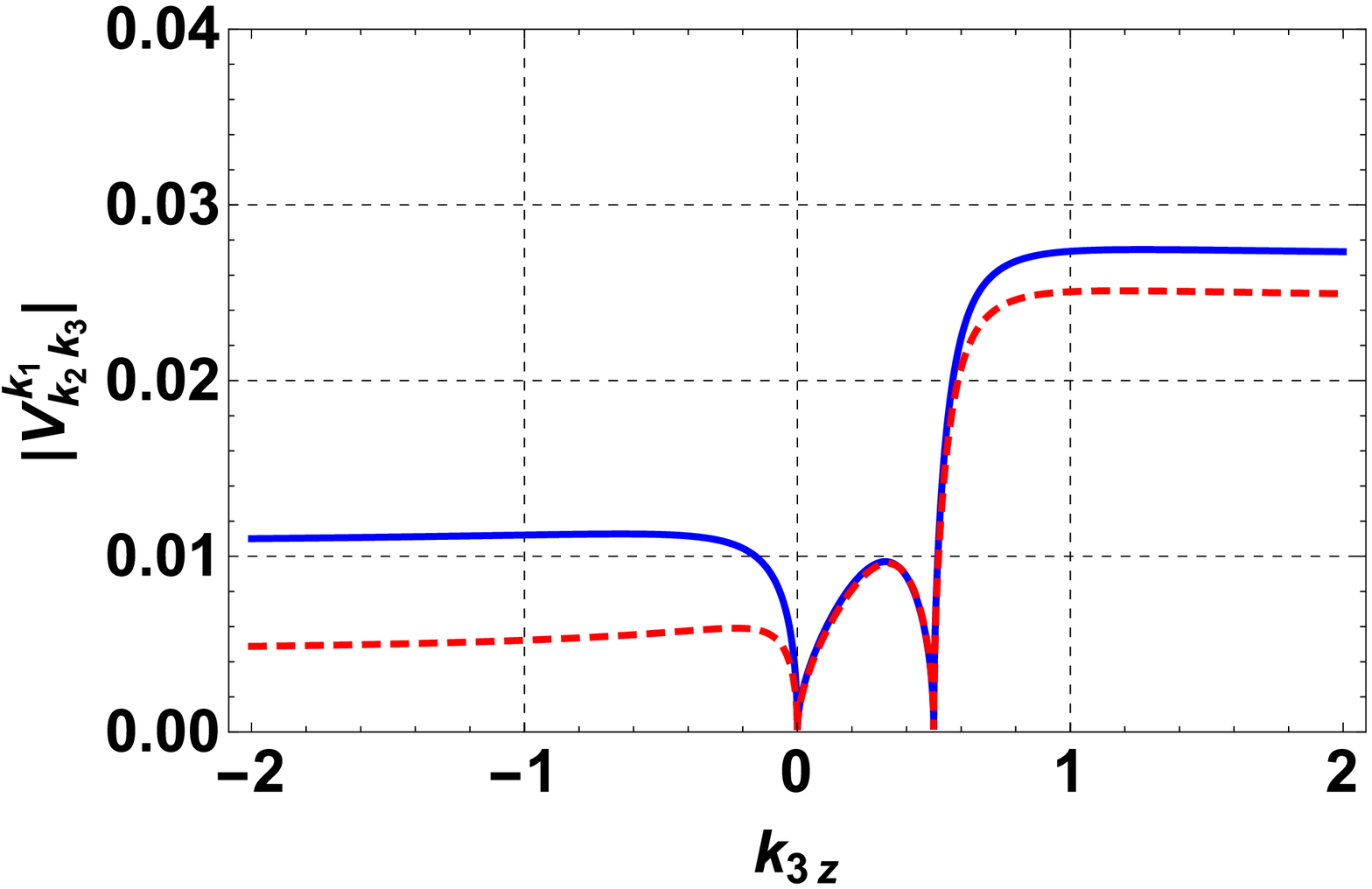}
\caption{\label{1Dsecond}
Comparison of the exact amplitude $|V^{\k_1}_{\k_2\k_3}|$ given by~\Ref{V123_p1} (blue solid lines) and its small-frequency asymptotics given by~\Ref{V123 asymptotic small omega} (red dashed lines). The resonance curves are presented for $\varphi_1=\frac{\pi}{4}$. LEFT: the primary wave vector $\k_1=\bigl(\sin\frac{7\pi}{16},0,\cos\frac{7\pi}{16}\bigr)$. RIGHT: $\k_1=\bigl(\sin\theta_p,0,\cos\theta_p\bigr)$.}
\end{figure}

The asymptotic~\Ref{V123 asymptotic small omega}, as a function of $\k_3$, is valid not only in the shortwave limit but for any point of the small-frequency resonance surface. As was found by~\cite{bayly1986three}, the addition of weak ellipticity to the circular flow of the fluid leads to excitation of parametric inertial waves with maximum increment at angles $\theta_p$, satisfying the condition $|\cos\theta_p|=\frac12$ (see also the \textit{Conclusion} section). In Fig.~\ref{1Dsecond} we demonstrate how the small-frequency asymptotic~\Ref{V123 asymptotic small omega} works at small frequencies and in the especially important case where $\theta = \theta_p = \frac{\pi}{3}$. We find that in the first case, the asymptotics~\Ref{V123 asymptotic small omega} perfectly describe the three-wave interaction processes on the whole resonance surface and can be used instead of the cumbersome exact expression~\Ref{V123_p1}. Meanwhile, for the latter case, the small frequency asymptotics~\Ref{V123 asymptotic small omega} work well only in the middle branch of the resonance surface.

The presented study of the instability increment is in agreement with results of~\cite{bordes2012experimental}, where the particular two-dimensional case (i.e., when $\k_1$, $\k_2$ and $\k_3$ lie in one plane) was studied experimentally and theoretically (using the helical modes decomposition approach). In particular, the nonzero values of the increment in the shortwave limit were also studied. Here, we provide a more detailed description of the increment in the whole $\k$-space, which we believe will be useful for the future experimental and numerical works devoted to the three-wave interactions of inertial waves.

\section{Turbulent cascade of the inertial waves}
Secondary waves, created as a result of the decay instability of the primary waves, are also unstable with respect to the same process~(\ref{decay law1}). Thus, one can expect that this multi-level decay instability will result in the formation of some type of energy cascade from the large scale toward the small scale, as in usual homogeneous turbulence. The energy spectrum in the inertial range (i.e., at intermediate scale) of the cascade can be found from the well-known kinetic equation, which for three-wave processes has the following form:
\begin{eqnarray}\label{kinetic}
\frac{\pt n(\k_1,t)}{\pt t} = \pi\int\biggl [ |V^{\k_1}_{\k_2\k_3}|^2 N^{\k_1}_{\k_2\k_3}\delta_{\k_1-\k_2-\k_3} \delta_{\omega_{\k_1}-\omega_{\k_2}-\omega_{\k_3}}
\\\nonumber
+2 |V^{\k_2}_{\k_1\k_3}|^2 N^{\k_2}_{\k_1\k_3}\delta_{\k_2-\k_1-\k_3} \delta_{\omega_{\k_2}-\omega_{\k_1}-\omega_{\k_3}} \biggr ] d\k_2 d\k_3\,.
\end{eqnarray}
Here
\begin{equation}
N^{\k_1}_{\k_2\k_3} = n_{\k_2}n_{\k_3}-n_{\k_1}(n_{\k_2}+n_{\k_3})\,,
\end{equation}
and $n_{\k}$ is the wave density in $\k$-space. The solution of the kinetic equation \REF{kinetic} can be found only in particular cases, when both matrix elements and the eigen-frequency have some scaling properties (see the book \cite{zakharov1992kolmogorov}). Unfortunately, this is not the case for inertial waves in rotating fluid.

However, it is well-established that the statistical behaviour of inertial waves in rapidly rotating fluids is highly anisotropic (see, for instance, \cite{galtier2003weak}, \cite{sen2012anisotropy} or the book of~\cite{nazarenko2011wave} and references therein).  The dispersion law~\Ref{dispersion} leads to the formation of an energy cascade predominantly in the $\k_{\perp}$ - direction, when the wave frequencies are small and our simple asymptotic (\ref{V123 asymptotic small omega}) is valid. Under this approximation, the dispersion law and interaction amplitude become scale-invariant with respect to scaling of $k_z$ and $k_{\perp}$:
\begin{equation}\label{scaling}
\omega_{\k} = 2\O\frac{k_z}{k_{\perp}}, \,\,\,\,\,\,\,\,\,\,\,\, V^{\k_2}_{\k_1\k_3} \sim \sqrt{\O k_z k_{\perp}}.
\end{equation}
Solutions of the kinetic equation in this case were analyzed in detail in \cite{zakharov1992kolmogorov}. The result is that for the scaling, given by \REF{scaling}, stationary solutions of the kinetic equation, corresponding to the constant densities of the energy flux $\mathcal P_1$ and momentum flux $\mathcal P_2$ along $\^\z$ direction can be found as:
\begin{eqnarray}\label{kz spectrum}
\mathcal E_1(\k) \equiv \mathcal P_1^{1/2}k_z^{-1/2}k_{\perp}^{-7/2},   \,\,\,\,\,\,\,\,\,\,\,\, \mathcal E_2(\k) \equiv \mathcal P_2^{1/2}k_z^{-1/2}k_{\perp}^{-4}.
\end{eqnarray}
These anisotropic weak turbulent energy spectra were first identified by~\cite{galtier2003weak} using the helical modes decomposition approach. Here, we present spectra not averaged by the azimuth angle $\varphi$, which explains the difference in the power laws of~\Ref{kz spectrum} and the paper of~\cite{galtier2003weak}.

\section{Conclusion}

In this work, we developed the complete Hamiltonian formalism for inertial waves in rotating fluid. Namely, we suggest Hamiltonian variables for the velocity field and found all terms of the exact Hamiltonian, which consist of linear (diagonal), three- and four-wave interaction parts. In the important case of rapid rotation, the four-wave interaction term has second-order accuracy by the Rossby number, so the dynamics of the inertial waves are mostly determined by the first-order three-wave interaction processes. Special attention was paid to the process of wave decay $\k_1 \ra \k_2+\k_3$, which is responsible for the decay instability and plays a central role in the dynamics of inertial waves and the formation of anisotropic wave turbulence spectra. We discussed the geometry of the resonance surface for three-wave interactions, which has a highly nontrivial form with shortwave infinite branches. Then, we studied the behaviour of the increased decay instability on the whole three-dimensional resonance surface. In addition, we calculated several key asymptotics of the three-wave matrix element to improve the understanding of the main characteristics of the decay instability, which is useful for future studies. We believe that the newly identifed features can be observed in numerical simulations and natural experiments, such as those presented in the works of \cite{bordes2012experimental} and \cite{di2016quantifying}.

The structure of the decay instability increment on the resonance surface can be straightforwardly observed in numerical simulations on a discrete grid (see, e.g., the work of~\cite{dyachenko2003decay} devoted to decay of capillary waves); meanwhile, such research of separate waves dynamics becomes nontrivial in natural experiments. The recently developed experimental method to study the dynamics of \textit{internal waves} in a stratified fluid is a promising tool that can also be applied to inertial waves~\citep{scolan2013nonlinear,brouzet2016inertial}. Indeed, this approach is based on the reflection (from the rigid boundary) properties of the internal waves, making it possible to focus them in a wave attractor area using a special trapezoid configuration of the experimental fluid tank~(\cite{maas1997observation}). The reflection properties of internal waves are determined by the following dispersion law: $\omega_{\k}=\omega_0 |\sin\theta_{\k}|$. Here, the constant $\omega_0$ is the so-called \textit{buoyancy frequency} and $\theta_{\k}$ is the angle between the wave vector and the direction of stratification~(see, e.g., \cite{phillips1977dynamics}). The dispersion law~\Ref{dispersion} determines the analogous wave reflection properties, enabling the creation of experimental wave attractors for inertial waves in rotating fluid (see also the discussion in ~\citep{brouzet2016inertial}).

Although internal and inertial waves have similar dispersion laws, their physical properties have significant differences. For example, consider the small-frequency limit, which plays the key role in the formation of the energy cascade for a rapidly rotating fluid (see the previous section), as well as for a strongly stratified fluid. The latter enables the existence of important nonpropagating modes escaping the wave regime. Thus, the turbulent behaviour of the strongly stratified fluid is defined by the coexistence of the internal wave motions and the nonpropagating modes (see chapters \textit{4} and \textit{7} in the monograph of~\cite{sagaut2008homogeneous}, where a detailed comparison of rapidly rotating and strongly stratified fluid was presented). In a rapidly rotating fluid, the two-dimensional nonpropagating modes appear only as a singular limit when $k_z = 0$, in accordance with Teylor-Proudman theorem, and can thus be excluded from wave turbulence theory. Moreover, the Hamiltonian formalism for a stratified fluid is based on another pair of Clebsch variables, the velocity potential and a special functional of the fluid density and the Montgomery potential, so the matrix elements have a different structure~(see the papers of \cite{lvov2001hamiltonian,lvov2004energy}).

Our study of the turbulent cascade confirmed the results previously obtained in the frame of helical modes decomposition~\citep{galtier2003weak,galtier2014theory}. However, the presented kinetic equation (\ref{kinetic}) enables further study. For example, the hypothesis of the local nature for the rotating turbulence should be verified by studying the convergence of the integral of collisions. Additionally, more general solutions of the kinetic equation (\ref{kinetic}) can be obtained numerically.

At first sight, in the case of rapid rotation, the four-wave interactions can be considered only to obtain more accurate predictions. However, such interactions play the central role in the so-called \textit{parametric wave turbulence}. The corresponding theory was introduced by V.E.~Zakharov, V.S.~L'vov and S.S.~Starobinets for spin waves in the paper of \cite{zakharov1971stationary} and was then developed in detail for general Hamiltonian systems (see the book of~\cite{Lvov1994wave}). Waves excited by pairs as a result of \textit{parametric instability} conserve their phase correlation in four-wave interaction processes. This leads to the special \textit{phase mechanism} of wave amplitude saturation in the turbulent state. As previously mentioned, for inertial waves, the role of \textit{parametric pumping} can result in weak ellipticity of the flow. The suggested Hamiltonian formalism for circular flow can be easily generalised to the weakly elliptic case, which enables the use of the presented results for the future development of parametric wave turbulence theory for inertial waves.



\section{Appendix}
Note that we use prime symbols here to distinguish the rotating reference frame from the inertial frame. In the main text of the manuscript, these primes are omitted.
\subsection{Hamiltonian description of fluid in the rotating reference frame}
The conventional Clebsch representation of the velocity field $\v$ for the Euler equations
\begin{subequations}\label{nonrotEuler}
\begin{eqnarray}\label{nonrotEuler-a}
 \pt_t \v + (\v\boldsymbol{\cdot}\n)\v &=& -\n p\,,
\\
\label{nonrotEuler-b}
 \n\boldsymbol{\cdot}\v &=& 0\,,
\end{eqnarray}
\end{subequations}
has the following form (see, e.g.~\cite{lamb1945hydrodynamics} or \cite{zakharov1992kolmogorov}):
\begin{eqnarray}\label{Clebsch-R nonr}
  \v = \lambda\bn\mu + \n\Phi\,.
\end{eqnarray}
Considering the continuity equation~\Ref{nonrotEuler-b}, we can rewrite~\Ref{Clebsch-R nonr} using the transverse projector~\Ref{projector} as
\begin{equation}\label{Clebsch-P nonr}
  \v = \^{\P}\*\left(
    \lambda\bn\mu
  \right)\,,
\end{equation}
(see also the second chapter in~\citep{sagaut2008homogeneous} concerning the use of projection operators). To find an analog of this representation for a rotating fluid (i.e., for~\REF{rotEuler}), we consider a change of frame as the following transformation of variables:
\begin{subequations}\label{rotation}
\begin{eqnarray}
\lambda(\r,t) \ra \lambda^{\prime}(\r,^{\prime}t) \equiv \lambda^{\prime}[t,\mathcal{\widehat{R}}(t)\cdot \r^{\prime}]\,,
\\
\mu(\r,t) \ra \mu^{\prime}(\r^{\prime},t) \equiv \mu^{\prime}[t,\mathcal{\widehat{R}}(t)\cdot \r^{\prime}]\,.
\end{eqnarray}
\end{subequations}
Here, $\r^{\prime}$ is the position vector in the new (rotating with angular velocity $\bO\, \| \, \^\z$) reference frame, while $\r = \mathcal{\widehat{R}}(t)\cdot \r^{\prime}$ corresponds to the initial inertial frame. The transformation matrix $\mathcal{\widehat{R}}(t)$ in our case is given by
\begin{equation}
  \mathcal{\widehat{R}}(t) = \(\begin{array}{ccc}
    \cos\Omega t & -\sin\Omega t & 0 \\
    \sin\Omega t & \cos\Omega t & 0 \\
    0 & 0 & 1 \\
  \end{array}\)\,.
\end{equation}
The transformation of variables \Ref{rotation} is not canonical: new fields $\lambda'$, $\mu'$ obey the equations of motion (see \REF{eq-H}) with a different Hamiltonian:
\begin{equation}\label{newHamiltonian}
\H^{\prime} \equiv \H - \int (\bO\times\r^{\prime})\*(\lambda^{\prime}\bn\mu^{\prime})d\r^{\prime} \,. 
\end{equation}
Here, the gradient $\bn\equiv\frac{\pt}{\pt \r^{\prime}}$ is defined for the rotating reference frame. The main Hamiltonian $\H$ in the new coordinates has the same form as in the inertial frame~\Ref{Hamiltonian}, which can be explained as follows. The Jacobian of the transformation $\r^{\prime} \ra \r$ is unity. $\det\mathcal{\widehat{R}}(t)=1$ and the operators $\bn$ and $\^{\P}$ are instantaneous, so their isotropic scalar combinations do not change under rotation.

The term $(\bO\times\r)$ in~\Ref{newHamiltonian} is divergent-less. Thus, by adding the constant $\frac12\int|(\bO\times\r)|^2d\r$ to the equation~\Ref{newHamiltonian}, we can write the Hamiltonian in the rotating frame as:
\begin{equation}\label{Hamiltonian-R2}
  \H^{\prime} = \int \frac12\left|\v^{\prime}\right|^2 d\r^{\prime}\,,
\end{equation}
where
\begin{equation}\label{Clebsch-R}
\v^{\prime} \equiv \^{\P}\*\(\lambda^{\prime}\bn\mu^{\prime}\) - (\bO\times\r^{\prime})\,.
\end{equation}
Again, the operators ($\^{\P}$ and $\bn$) act on the variable $\r^{\prime}$. The new field $\v^{\prime}$ in the~\REF{Clebsch-R} is simply the velocity of fluid in the rotating frame; thus, the only difference between Hamiltonian approaches in the inertial and rotating frames is the form of the Clebsch transformation -- compare equations~\Ref{Clebsch-P nonr} and~\Ref{Clebsch-R}.

\subsection{Clebsch representation of the velocity field in the rotating reference frame}
First, we consider the fluid rotating as a whole with angular velocity $\bO\, \| \,\^\z$. In the inertial frame, such \textit{basic} circular flow $\v_0$ can be presented as:
\begin{equation}\label{basic flow}
\v_0 = \bO\times\r = -\^\x\O y + \^\y\O x\,.
\end{equation}
The Clebsch fields $\lambda_0$ and $\mu_0$ for the basic flow $\v_0$ and the corresponding potential $\Phi_0$ satisfying equation
\begin{eqnarray}
  \v_0 = \lambda_0\bn\mu_0 + \n\Phi_0\,,
\end{eqnarray}
can be found in different forms (see the discussion in section (3.1)). We suggest the simplest form, which can be obtained for the linear dependence of $\lambda_0$, $\mu_0$ on coordinates $\r$:
\begin{subequations}\label{calibration2}
\begin{eqnarray}
\lambda_0 = \brho_{\lambda} \cdot \r = \rho_{\lambda x}x + \rho_{\lambda y}y \,,
\\
\mu_0 = \brho_{\mu} \cdot \r = \rho_{\mu x}x + \rho_{\mu y}y \,.
\end{eqnarray}
\end{subequations}
Here, $\brho_{\lambda,\mu}=\rho_{\lambda,\mu x}\^\x + \rho_{\lambda,\mu x}\^\y$ are time-dependent vectors. Then, for the pontial
$\Phi_0$, we obatin a couple of equations:
\begin{subequations}\label{Phi_0_Eq1}
\begin{eqnarray}
\pt_x \Phi_0 = -\rho_{\mu x} \brho_{\lambda}\cdot\r\,,
\\
\pt_y \Phi_0 = -\rho_{\mu y} \brho_{\lambda}\cdot\r\,.
\end{eqnarray}
\end{subequations}
The compatibility condition for~\Ref{Phi_0_Eq1} $\pt_x \pt_y \Phi_0 = \pt_y \pt_x \Phi_0$ gives only the one restriction to the possible choice of vectors $\brho_{\lambda,\mu}$:
\begin{equation}\label{restriction}
\rho_{\lambda x} \rho_{\mu y} - \rho_{\lambda y} \rho_{\mu x} = 2\O \,.
\end{equation}
Then, the potential $\Phi_0$ can be found as
\begin{equation}
\Phi_0 = -\frac12 \lambda_0\mu_0\,.
\end{equation}
Note that the left-hand side of \REF{restriction} is simply the determinant of the transformation~\Ref{calibration2} from $x,y$  to $\lambda,\mu$. Thus, the transformation~\Ref{calibration2} is always non-degenerate and can be inverted as:
\begin{subequations}
\begin{eqnarray}
x = \frac{1}{2\O}\bigl(\rho_{\mu y} \lambda_0 - \rho_{\lambda y} \mu_0 \bigr)\,,
\\
y = \frac{1}{2\O}\bigl(-\rho_{\mu x} \lambda_0 + \rho_{\lambda x} \mu_0 \bigr)\,.
\end{eqnarray}
\end{subequations}
It is convenient to choose $\lambda_0$ and $\mu_0$ as time-dependent rotating vectors, which will be stationary in the new rotating frame:
\begin{subequations}
\begin{eqnarray}
 \brho_{\lambda} =  \sqrt{2\O} (\^\x \cos\O t + \^\y \sin\O t)\,,
 \\
 \brho_{\mu}  =  \sqrt{2\O} (-\^\x \sin\O t + \^\y \cos\O t)\,.
\end{eqnarray}
\end{subequations}
$\brho_{\lambda}$ and $\brho_{\mu}$ are directed along $\^\x^{\prime}$ and $\^\y^{\prime}$, respectively. Then, in the rotating reference frame, the Clebsch variables for basic flow~\Ref{basic flow} are given by:
\begin{equation}\label{basic Clebsch}
 \lambda^{\prime}_0 = \sqrt{2\O} x^{\prime}, \,\,\,\,\,\,\,\,\,\, \mu^{\prime}_0 = \sqrt{2\O} y^{\prime}\,.
\end{equation}

In this work, we study only the weak velocity perturbations on the background of the basic circular flow~\Ref{basic flow}. Thus, it is convenient to introduce deviations
$\tilde{\lambda}^{\prime} \equiv \lambda^{\prime} - \lambda_0^{\prime}$ and $\tilde{\mu}^{\prime} \equiv \mu^{\prime} - \mu_0^{\prime}$ of the real flow from the basic flow. In the rotating reference frame, the basic flow velocity $\v_0^{\prime}$ is obviously zero; however, it is useful to formally consider the expression for velocity perturbation $\tilde{\v}^{\prime} \equiv \v^{\prime} - \v_0^{\prime}$ and to verify that 
\begin{equation}\label{dv pertr}
\tilde{\v}^{\prime} = \^{\P}\*\(\lambda^{\prime}\bn\mu^{\prime}\) - \^{\P}\*\(\lambda_0^{\prime}\bn\mu_0^{\prime}\) =
 \^{\P}\*\left[\sqrt{2\Omega}\(\tilde{\lambda}^{\prime}\^\y^{\prime}-\tilde{\mu}^{\prime}\^\x^{\prime}\) +\tilde{\lambda}^{\prime}\bn\tilde{\mu}^{\prime} \right]\,.
\end{equation}
The last equality in the expression~\Ref{dv pertr} can be verified using \REF{basic Clebsch} and the property of the projector operator to cancel the full gradient of an arbitrary function. 
The canonically-conjugated fields $\tilde{\lambda}^{\prime}$ and $\tilde{\mu}^{\prime}$ obey the usual equations of motion \Ref{eq-H}. Thus, omitting all the primes and tildes, we prove the velocity representations~\Ref{dv1} and \Ref{dv2}, which we introduced at the beginning of the paper.

\subsection{Coefficients of the interaction Hamiltonian}
Here, we present exact expressions for the amplitudes of the interaction part of the Hamiltonian $\H_{int}$, which were discussed in the section~(\ref{section interaction}). First, we write the amplitude for the three-wave process $\k_1+\k_2+\k_3 \ra 0$:
\begin{small}
\begin{eqnarray}\label{U123}
 U_{\k_1\k_2\k_3} = -\frac{3 i \sqrt{2\Omega}}{(2\upi)^{3/2}\kp\kpp\kppp}
\biggl\{
-\sqrt{\frac{\Omega^3}{\omega_{\k_1}\omega_{\k_2}\omega_{\k_3}}} F^{\k_1}_{\k_2\k_3} S_{\k_3\k_2}+ \frac{i}{8} \sqrt{\frac{\omega_{\k_1}\omega_{\k_2}\omega_{\k_3}}{\Omega^3}}S_{\k_3\k_2}
\,\,\,\,\,\,\,\,\,\,\,\,\,\,\,\,\,\,\,\,\,\,\,\,\,\,\,\
\\\nonumber
+\frac14 \sqrt{\frac{\omega_{\k_2}\omega_{\k_3}}{\O \, \omega_{\k_1}}} F^{\k_1}_{\k_2\k_3}S_{\k_3\k_2}-\frac{i}{2}\sqrt{\frac{\O \, \omega_{\k_1}}{\omega_{\k_2}\omega_{\k_3}}} S_{\k_3\k_2}^2-\frac{i}{2}\sqrt{\frac{\O \, \omega_{\k_3}}{\omega_{\k_1}\omega_{\k_2}}}F^{\k_1}_{\k_2\k_3}(\bkpp\*\bkppp)
\\\nonumber
+\frac{i}{2}\sqrt{\frac{\O \, \omega_{\k_2}}{\omega_{\k_1}\omega_{\k_3}}}F^{\k_1}_{\k_2\k_3}(\bkpp\*\bkppp)
+\frac{1}{4}\sqrt{\frac{\omega_{\k_1}\omega_{\k_3}}{\O \, \omega_{\k_2}}}S_{\k_3\k_2}(\bkpp\*\bkppp) 
-\frac{1}{4}\sqrt{\frac{\omega_{\k_1}\omega_{\k_2}}{\O \, \omega_{\k_3}}}S_{\k_3\k_2}(\bkpp\*\bkppp)
\biggr\}\,.
\end{eqnarray}
\end{small}
Then, we present amplitudes for all possible four-wave processes, such as $\k_1+\k_2 + \k_3 \ra \k_4$:
\begin{small}
\begin{eqnarray}
G^{\k_2\k_3\k_4}_{\k_1} =  \frac{1}{(2\upi)^3 \kp\kpp\kppp\kpppp} \biggl\{
\,\,\,\,\,\,\,\,\,\,\,\,\,\,\,\,\,\,\,\,\,\,\,\,\,\,\,\,\,\,\,\,\,\,\,\,\,\,\,\,\,\,\,\,\,\,\,\,\,\,\,\,\,\,\,\,\,\,\,\,\,\,\,\,\,\,\,\,\,\,\,\,\,\,\,\,\,\,\,\,\,\,\,\,\,\,\,\,\,\,\,\,\,\,\,\,\,\,\,\,\,\,\,\,\,\,\,\,\,\,\,\,\,\,\,\,\,\,\,\,
\,\,\,\,\,\,\,\,\,\,\,\,\,\,\,\,\,\,\,
\\\nonumber
\biggl(\sqrt{\frac{\Omega^4}{\omega_{\k_1}\omega_{\k_2}\omega_{\k_3}\omega_{\k_4}}} - \frac{1}{16}\sqrt{\frac{\omega_{\k_1}\omega_{\k_2}\omega_{\k_3}\omega_{\k_4}}{\Omega^4}}\biggr)
\biggl(S_{\k_1\k_4}S_{\k_3\k_2}Q^{\k_3,\k_2}_{-\k_1,\k_4} + S_{\k_1\k_3}S_{\k_2\k_4}Q^{\k_4,\k_2}_{\k_3,-\k_1}\biggr)
\\\nonumber
-\frac14\biggl(\sqrt{\frac{\omega_{\k_1}\omega_{\k_2}}{\omega_{\k_3}\omega_{\k_4}}} -
\sqrt{\frac{\omega_{\k_3}\omega_{\k_4}}{\omega_{\k_1}\omega_{\k_2}}}\biggr)
\biggl((\bkp\*\bkpppp)(\bkpp\*\bkppp)Q^{\k_3,\k_2}_{-\k_1,\k_4} - (\bkp\*\bkppp)(\bkpp\*\bkpppp) Q^{\k_4,\k_2}_{\k_3,-\k_1}\biggr)
\\\nonumber
+\frac14\biggl(\sqrt{\frac{\omega_{\k_1}\omega_{\k_3}}{\omega_{\k_2}\omega_{\k_4}}} -
\sqrt{\frac{\omega_{\k_2}\omega_{\k_4}}{\omega_{\k_1}\omega_{\k_3}}}\biggr)
\biggl((\bkp\*\bkpppp)(\bkpp\*\bkppp)Q^{\k_3,\k_2}_{-\k_1,\k_4} + S_{\k_1\k_3}S_{\k_2\k_4} Q^{\k_4,\k_2}_{\k_3,-\k_1}\biggr)
\\\nonumber
+\frac14\biggl(\sqrt{\frac{\omega_{\k_1}\omega_{\k_4}}{\omega_{\k_2}\omega_{\k_3}}} -
\sqrt{\frac{\omega_{\k_2}\omega_{\k_3}}{\omega_{\k_1}\omega_{\k_4}}}\biggr)
\bigl[S_{\k_1\k_4}S_{\k_3\k_2} Q^{\k_3,\k_2}_{-\k_1,\k_4} - (\bkp\*\bkppp)(\bkpp\*\bkpppp)Q^{\k_4,\k_2}_{\k_3,-\k_1}\biggr)+
\\\nonumber
+\frac{i}{2}\biggl(\sqrt{\frac{\Omega^2\omega_{\k_1}}{\omega_{\k_2}\omega_{\k_3}\omega_{\k_4}}}
 -\frac14\sqrt{\frac{\omega_{\k_2}\omega_{\k_3}\omega_{\k_4}}{\Omega^2\omega_{\k_1}}}\biggr)
 \bigl[S_{\k_3\k_2} (\bkp\*\bkpppp) Q^{\k_3,\k_2}_{-\k_1,\k_4} - S_{\k_4\k_2}(\bkp\*\bkppp)Q^{\k_4,\k_2}_{\k_3,-\k_1} \bigr]
\\\nonumber
+\frac{i}{2}\biggl(\sqrt{\frac{\Omega^2\omega_{\k_2}}{\omega_{\k_1}\omega_{\k_3}\omega_{\k_4}}}
 -\frac14\sqrt{\frac{\omega_{\k_1}\omega_{\k_3}\omega_{\k_4}}{\Omega^2\omega_{\k_2}}}\biggr)
 \bigl[S_{\k_1\k_4} (\bkpp\*\bkppp) Q^{\k_3,\k_2}_{-\k_1,\k_4} + S_{\k_3\k_1}(\bkpp\*\bkpppp)Q^{\k_4,\k_2}_{\k_3,-\k_1} \bigr]
 \\\nonumber
 -\frac{i}{2}\biggl(\sqrt{\frac{\Omega^2\omega_{\k_3}}{\omega_{\k_1}\omega_{\k_2}\omega_{\k_4}}}
 -\frac14\sqrt{\frac{\omega_{\k_1}\omega_{\k_2}\omega_{\k_4}}{\Omega^2\omega_{\k_3}}}\biggr)
 \bigl[S_{\k_1\k_4} (\bkpp\*\bkppp) Q^{\k_3,\k_2}_{-\k_1,\k_4} + S_{\k_4\k_2}(\bkp\*\bkppp)Q^{\k_4,\k_2}_{\k_3,-\k_1} \bigr]
 \\\nonumber
  +\frac{i}{2}\biggl(\sqrt{\frac{\Omega^2\omega_{\k_4}}{\omega_{\k_1}\omega_{\k_2}\omega_{\k_3}}}
 -\frac14\sqrt{\frac{\omega_{\k_1}\omega_{\k_2}\omega_{\k_3}}{\Omega^2\omega_{\k_4}}}\biggr)
 \bigl[S_{\k_3\k_2} (\bkp\*\bkpppp) Q^{\k_3,\k_2}_{-\k_1,\k_4} - S_{\k_3\k_1}(\bkpp\*\bkpppp)Q^{\k_4,\k_2}_{\k_3,-\k_1} \bigr]\biggr\} \,;
\end{eqnarray}
\end{small}
then $\k_1+\k_2+\k_3+\k_4 \ra 0$:
\begin{small}
\begin{eqnarray}
R_{\k_1\k_2\k_3\k_4} = -\frac{1}{(2\upi)^3}\frac{Q^{\k_1\k_2}_{\k_3\k_4}}{{\kp\kpp\kppp\kpppp}} \biggl\{
\,\,\,\,\,\,\,\,\,\,\,\,\,\,\,\,\,\,\,\,\,\,\,\,\,\,\,\,\,\,\,\,\,\,\,\,\,\,\,\,\,\,\,\,\,\,\,\,\,\,\,\,\,\,\,\,\,\,\,\,\,\,\,\,\,\,\,\,\,\,\,\,\,\,\,\,\,\,\,\,\,\,\,\,\,\,\,\,\,\,\,\,\,\,\,\,\,\,\,\,\,\,\,\,\,\,\,\,\,\,\,\,\,\,\,\,\,\,\,\,\,\,\,\,\,\,
\,\,\,\,\,\,\,\,\,\,\,\,\,\,\,
\\
\nonumber
\biggl(\sqrt{\frac{\Omega^4}{\omega_{\k_1}\omega_{\k_2}\omega_{\k_3}\omega_{\k_4}}}
+ \frac{1}{16}\sqrt{\frac{\omega_{\k_1}\omega_{\k_2}\omega_{\k_3}\omega_{\k_4}}{\Omega^4}}
-\frac14\sqrt{\frac{\omega_{\k_1}\omega_{\k_2}}{\omega_{\k_3}\omega_{\k_4}}}
-\frac14\sqrt{\frac{\omega_{\k_3}\omega_{\k_4}}{\omega_{\k_1}\omega_{\k_2}}}
\biggr)S_{\k_1\k_2}S_{\k_3\k_4} 
\\
\nonumber
+ \frac14 \biggl(\sqrt{\frac{\omega_{\k_1}\omega_{\k_4}}{\omega_{\k_2}\omega_{\k_3}}}
+\sqrt{\frac{\omega_{\k_2}\omega_{\k_3}}{\omega_{\k_1}\omega_{\k_4}}}
-\sqrt{\frac{\omega_{\k_1}\omega_{\k_3}}{\omega_{\k_2}\omega_{\k_4}}}
-\sqrt{\frac{\omega_{\k_2}\omega_{\k_4}}{\omega_{\k_1}\omega_{\k_3}}}
\biggr) (\bkp\*\bkpp)(\bkppp\*\bkpppp)
\\
\nonumber
+ \frac{i}{2}\biggl(
\sqrt{\frac{\Omega^2\omega_{\k_3}}{\omega_{\k_1}\omega_{\k_2}\omega_{\k_4}}}
-\sqrt{\frac{\Omega^2\omega_{\k_4}}{\omega_{\k_1}\omega_{\k_2}\omega_{\k_3}}}
+\frac14\sqrt{\frac{\omega_{\k_1}\omega_{\k_2}\omega_{\k_4}}{\Omega^2\omega_{\k_3}}}
-\frac14\sqrt{\frac{\omega_{\k_1}\omega_{\k_2}\omega_{\k_3}}{\Omega^2\omega_{\k_4}}}
\biggr)S_{\k_1\k_2} (\bkppp\*\bkpppp)
\\
\nonumber
+ \frac{i}{2} \biggl(\sqrt{\frac{\Omega^2\omega_{\k_1}}{\omega_{\k_2}\omega_{\k_3}\omega_{\k_4}}}
-\sqrt{\frac{\Omega^2\omega_{\k_2}}{\omega_{\k_1}\omega_{\k_3}\omega_{\k_4}}}
+\frac14\sqrt{\frac{\omega_{\k_2}\omega_{\k_3}\omega_{\k_4}}{\Omega^2\omega_{\k_1}}}
-\frac14\sqrt{\frac{\omega_{\k_1}\omega_{\k_3}\omega_{\k_4}}{\Omega^2\omega_{\k_2}}}
\biggr)S_{\k_3\k_4} (\bkp\*\bkpp) \biggr\}\,;
\end{eqnarray}
\end{small}
and finally for the process $\k_1+\k_2 \ra \k_3+\k_4$:
\begin{small}
\begin{eqnarray}
W^{\k_1\k_2}_{\k_3\k_4}  =  \frac{1}{(2\upi)^3 \kp\kpp\kppp\kpppp} \biggl\{
\biggl(\sqrt{\frac{\Omega^4}{\omega_{\k_1}\omega_{\k_2}\omega_{\k_3}\omega_{\k_4}}} + \frac{1}{16}\sqrt{\frac{\omega_{\k_1}\omega_{\k_2}\omega_{\k_3}\omega_{\k_4}}{\Omega^4}}\biggr)\times \,\,\,\,\,\,\,\,\,\,\,\,\,\,\,\,\,\,\,\,\,\,\,\,\,\,\,\
\\\nonumber
\times\bigl[S_{\k_1\k_2}S_{\k_3\k_4}Q^{\k_1,\k_2}_{\k_3,\k_4} + S_{\k_1\k_4}S_{\k_3\k_2}Q^{\k_1,-\k_4}_{\k_3,-\k_2} + S_{\k_1\k_3}S_{\k_2\k_4}Q^{\k_1,-\k_3}_{-\k_2,\k_4}\bigr]-
\\\nonumber
-\frac14\biggl(\sqrt{\frac{\omega_{\k_1}\omega_{\k_2}}{\omega_{\k_3}\omega_{\k_4}}} +
\sqrt{\frac{\omega_{\k_3}\omega_{\k_4}}{\omega_{\k_1}\omega_{\k_2}}}\biggr)\times
\\\nonumber
\times\bigl[S_{\k_1\k_2}S_{\k_3\k_4}Q^{\k_1,\k_2}_{\k_3,\k_4} - (\bkp\*\bkpppp)(\bkpp\*\bkppp)Q^{\k_1,-\k_4}_{\k_3,-\k_2} + (\bkp\*\bkppp)(\bkpp\*\bkpppp) Q^{\k_1,-\k_3}_{-\k_2,\k_4}\bigr]
\\\nonumber
+\frac14\biggl(\sqrt{\frac{\omega_{\k_1}\omega_{\k_3}}{\omega_{\k_2}\omega_{\k_4}}} +
\sqrt{\frac{\omega_{\k_2}\omega_{\k_4}}{\omega_{\k_1}\omega_{\k_3}}}\biggr)\times
\\\nonumber
\times\bigl[(\bkp\*\bkpp)(\bkppp\*\bkpppp) Q^{\k_1,\k_2}_{\k_3,\k_4} + (\bkp\*\bkpppp)(\bkpp\*\bkppp)Q^{\k_1,-\k_4}_{\k_3,-\k_2} + S_{\k_1\k_3}S_{\k_2\k_4} Q^{\k_1,-\k_3}_{-\k_2,\k_4}\bigr] -
\\\nonumber
-\frac14\biggl(\sqrt{\frac{\omega_{\k_1}\omega_{\k_4}}{\omega_{\k_2}\omega_{\k_3}}} +
\sqrt{\frac{\omega_{\k_2}\omega_{\k_3}}{\omega_{\k_1}\omega_{\k_4}}}\biggr)\times
\\\nonumber
\times\bigl[(\bkp\*\bkpp)(\bkppp\*\bkpppp) Q^{\k_1,\k_2}_{\k_3,\k_4} - S_{\k_1\k_4}S_{\k_3\k_2}Q^{\k_1,-\k_4}_{\k_3,-\k_2} + (\bkp\*\bkppp)(\bkpp\*\bkpppp)Q^{\k_1,-\k_3}_{-\k_2,\k_4}\bigr]+
\\\nonumber
+\frac{i}{2}\biggl(\sqrt{\frac{\Omega^2\omega_{\k_1}}{\omega_{\k_2}\omega_{\k_3}\omega_{\k_4}}}
 +\frac14\sqrt{\frac{\omega_{\k_2}\omega_{\k_3}\omega_{\k_4}}{\Omega^2\omega_{\k_1}}}\biggr)\times
 \\\nonumber
\times\bigl[S_{\k_3\k_4} (\bkp\*\bkpp) Q^{\k_1,\k_2}_{\k_3,\k_4} + S_{\k_3\k_2}(\bkp\*\bkpppp)Q^{\k_1,-\k_4}_{\k_3,-\k_2} \bigr]-
\\\nonumber
-\frac{i}{2}\biggl(\sqrt{\frac{\Omega^2\omega_{\k_2}}{\omega_{\k_1}\omega_{\k_3}\omega_{\k_4}}}
 +\frac14\sqrt{\frac{\omega_{\k_1}\omega_{\k_3}\omega_{\k_4}}{\Omega^2\omega_{\k_2}}}\biggr)\times
 \\\nonumber
 \times\bigl[S_{\k_3\k_4} (\bkp\*\bkpp) Q^{\k_1,\k_2}_{\k_3,\k_4} + S_{\k_1\k_4}(\bkpp\*\bkppp)Q^{\k_1,-\k_4}_{\k_3,-\k_2} \bigr]-
 \\\nonumber
 -\frac{i}{2}\biggl(\sqrt{\frac{\Omega^2\omega_{\k_3}}{\omega_{\k_1}\omega_{\k_2}\omega_{\k_4}}}
 +\frac14\sqrt{\frac{\omega_{\k_1}\omega_{\k_2}\omega_{\k_4}}{\Omega^2\omega_{\k_3}}}\biggr)\times
 \\\nonumber
\times\bigl[S_{\k_1\k_2} (\bkppp\*\bkpppp) Q^{\k_1,\k_2}_{\k_3,\k_4} + S_{\k_1\k_4}(\bkpp\*\bkppp)Q^{\k_1,-\k_4}_{\k_3,-\k_2} \bigr]+
 \\\nonumber
 +\frac{i}{2}\biggl(\sqrt{\frac{\Omega^2\omega_{\k_4}}{\omega_{\k_1}\omega_{\k_2}\omega_{\k_3}}}
 +\frac14\sqrt{\frac{\omega_{\k_1}\omega_{\k_2}\omega_{\k_3}}{\Omega^2\omega_{\k_4}}}\biggr)\times
 \\\nonumber
 \times\bigl[S_{\k_1\k_2} (\bkppp\*\bkpppp) Q^{\k_1,\k_2}_{\k_3,\k_4} + S_{\k_3\k_2}(\bkp\*\bkpppp)Q^{\k_1,-\k_4}_{\k_3,-\k_2} \bigr]\biggr\}\,.
\end{eqnarray}
\end{small}

\section{Acknowledgments} The work in sections (4) and (5) was performed with support from the Russian Science Foundation (Grand No. 14-22-00174). AG is also grateful for the support of the RFBR (Grant No. 16-31-60086 mol\_a\_dk). The authors thank the reviewers for providing constructive comments and suggestions on earlier versions of the manuscript.


\end{document}